\documentclass[11pt]{article}
\usepackage{amsmath}
\usepackage{amssymb}
\usepackage{amsfonts,amsthm,bm}
\usepackage{color,xcolor}
\usepackage[greek,english]{babel}

\usepackage{graphicx,epsfig}
\graphicspath{{figures/}{./}}
\usepackage{float}
\usepackage{subfigure}
\usepackage{tikz}

\usepackage[pdfencoding=auto,colorlinks=true,
allcolors=blue]{hyperref}% links
\newcommand{\be}{\begin{eqnarray}}
\newcommand{\ee}{\end{eqnarray}}
\newcommand{\bea}{\begin{eqnarray}}

\newcommand{\eea}{\end{eqnarray}}

\newcommand{\beq}{\begin{equation}}
\newcommand{\eeq}{\end{equation}}
\newcommand{\bseq}{\begin{subequations}}
\newcommand{\eseq}{\end{subequations}}

\title{\textbf{{{Echoes of Compact Objects in Scalar-Tensor Theories of Gravity }}}}

\author{Christoforos Vlachos\thanks{E-mail:cvlach@mail.ntua.gr},\,\,\,\, Eleftherios Papantonopoulos\thanks{E-mail:lpapa@central.ntua.gr}
\\
\textit{Physics Division, National Technical University of
Athens},
\\
\textit{15780 Zografou Campus, Athens, Greece}
\\
Kyriakos Destounis\thanks{E-mail:kyriakos.destounis@uni-tuebingen.de}
\\
\textit{Theoretical Astrophysics, IAAT, University of T$\ddot{u}$bingen, 72076 T$\ddot{u}$bingen, Germany}}

\date{}
\begin{document}
\large
\maketitle

\begin{abstract}
\noindent{Scalar-tensor theory predicts solutions to the gravitational field equations which describe compact objects in the presence of a non-minimally coupled scalar field to the Einstein tensor. These objects are black holes with scalar hair and wormholes supporting scalar phantom matter. The evolution of test fields in fixed asymptotically-flat backgrounds of exotic compact objects leads to the formation of echoes in the ringdown signal, which designate the existence of trapping regions close to the event horizon. Here, we consider minimally-coupled test scalar fields propagating on compact object solutions of the Horndeski action, which possess an effective cosmological constant, leading to anti-de Sitter asymptotics, and show that echoes can form in the ringdown waveform due to the entrapment of test fields between the photon sphere and the effective asymptotic boundary. Although the presence of an event horizon leads to the usual echoes with decaying amplitude, signifying modal stability of the scalarized black hole considered, we find that test scalar fields propagating on a scalarized wormhole solution give rise to echoes of constant and equal amplitude to that of the initial ringdown, indicating the existence of normal modes. Finally, we find that, near extremality, the test field exhibits a concatenation of echoes; the primary ones are associated with the trapping region between the photon sphere and the effective anti-de Sitter boundary while the secondary ones are linked to the existence of a potential well at the throat of the wormhole.}
\end{abstract}
\newpage
\tableofcontents

\section{Introduction}

The direct observation of gravitational waves (GWs) produced during the relativistic collision of two compact objects, offers exciting new opportunities for the study of the nature of the colliding bodies. In the near future, following the recent LIGO detections~\cite{Abbott:2016blz}-\cite{TheLIGOScientific:2017qsa}, GW astronomy will provide us a new understanding of the gravitational interaction and astrophysics in extreme-gravity conditions. The recent observations do not yet probe the detailed structure of spacetime beyond the photon sphere, however one expects that the strong gravity regime will in the next years  come to our understanding   with future GW observations. In particular, the expectation   is to precisely detect the ringdown phase, which is governed by a series of damped oscillatory modes at early times, named quasinormal modes (QNMs) \cite{Vishveshwara:1970zz}-\cite{Konoplya:2011qq}, and may potentially contain unexpected  anomalies due to new physics at late times \cite{Cardoso:2016rao}.

The expectation is that future GW observations will give us some information on the nature and physics of the near-horizon region of black holes (BHs) and if these regions exhibit any unexpected structure. Alternatives to BHs, that is, objects without event horizons, were recently constructed, known as exotic compact objects (ECOs) \cite{Mazur:2001fv}-\cite{Holdom:2016nek}. The existence of any structure at near-horizon scales would generate a series of ``echoes'' of the primary gravitational wave signal, produced during the ringdown phase \cite{Cardoso:2016rao,Cardoso:2016oxy}. The LIGO data have already been analyzed on the presence of echoes \cite{Abedi:2016hgu,Ashton:2016xff,Abedi:2017isz}.

It is  believed that the ringdown waveform is dominated by the QNMs
of the compact object remnant. Thus, the detection of overtones from the ringdown signal allows for precision measurements of the characteristic parameters of compact objects like the mass, charge and angular momentum. Various studies suggest that the ringdown signal is dominated by the mode excitations of photons trapped in unstable circular orbits at the photon sphere, namely the photon sphere (PS) modes \cite{Cardoso:2008bp}-\cite{Destounis:2020yav}. These QNMs are directly related to the existence of the PS, and if the compact object is an asymptotically flat BH no other oscillatory mode is excited. For ECOs, on the other hand, although the PS excitations still exist at the early stage of the ringdown signal, as in the case of  BHs, they do not belong to the QNM spectrum \cite{Cardoso:2016rao,Cardoso:2016oxy}.

Wormholes are ECO solutions of the Einstein equations that connect different parts of the Universe or two different Universes \cite{MisWheel,Wheel}. Although wormholes have distinct causal strutures from BHs, they possess PSs and, therefore, can disguise themselves as BHs in GW data if one only focuses on the early stage of the ringdown signal. Lorentzian wormholes in General Relativity (GR) were discussed in \cite{Morris,phantom,Visser}, where a static spherically-symmetric metric was introduced and conditions for traversable wormholes were found. Unfortunately, wormhole solutions of the Einstein equations lead to the violation of null energy condition (NEC). A matter distribution of exotic or phantom matter allows the formation of traversable wormhole geometries in GR. There have been many efforts to build a wormhole with ordinary matter satisfying the NEC \cite{Visser,Antoniou:2019awm,Mehdizadeh:2015dta} in modified gravity theories like Brans-Dicke theory \cite{Brans}, $f({\mathcal R})$ gravity \cite{FR}, Einstein-Gauss-Bonnet theory~\cite{GB}, Einstein-Cartan theory and general scalar-tensor theories \cite{Cartan}.

A series of contemporary studies \cite{Cardoso:2016rao,Cardoso:2016oxy} suggest that the ringdown signal may provide a conclusive proof for the formation of an event horizon. Such expectation is based on the assumption that an ECO would possess a reflective surface beyond the PS instead of an event horizon. This would lead to the existence of a trapping region between the surface of the ECO and PS where perturbations could be confined and manifest themselves as echoes in the late stage of the gravitational waveform. Then, by considering ringdown waveforms of ECOs, such as wormholes, it has been claimed \cite{Cardoso:2016rao,Cardoso:2016oxy} that precision observations of the late-time ringdown signal can distinguish between the formation of ECOs or BHs.

In this work we will consider the ringdown phase of exact compact object solutions of scalar-tensor theory, which is part of the Horndeski class of solutions \cite{horny} that give rise to second order field equations in four dimensions \cite{Nicolis:2008in,Deffayet:2009wt,Deffayet:2009mn} (for a review of this class of Horndeski theories see \cite{Papantonopoulos:2019eff}). The BH \cite{Kolyvaris:2011fk}-\cite{Charmousis:2014zaa} and wormhole \cite{Korolev:2014hwa} solutions we consider arise from a gravitational action with a real or phantom scalar field, respectively, non-minimally coupled to the Einstein tensor. These exact solutions encode the `gravitational' scalar, and its coupling strength, in the metric tensor components as a primary charge \cite{Rinaldi:2012vy,Korolev:2014hwa}, which asymptotically plays the role of an effective negative cosmological constant. The motivation for considering these objects is that they have a natural asymptotic reflective boundary, in the external region of the PS, which may lead to different behaviour compared with the case of flat spacetimes. The anti-de Sitter (AdS) spacetime is a necessity for holographic theories which were built by applying the gauge/gravity duality. The aim of holography is to study strongly coupled phenomena using dual gravitational systems where the coupling is weak \cite{Maldacena:1997re}. This duality, which is well founded in string theory, has many interesting applications and among them is condensed matter physics. In these theories, AdS BHs play an essential role in the gravity sector in order to achieve the abundant phase structure of the condensed matter system lying on the conformal boundary (for a review, see \cite{Hartnoll:2009sz}). These holographic theories stimulated the extended study of AdS BHs, their formation and their stability, with their QNMs describing the approach to thermal equilibrium in the dual conformal field theory on the boundary \cite{Horowitz:1999jd}.

Wormholes in AdS spacetimes where discussed in \cite{Maldacena:2004rf}, in an attempt to yield some information about the physics of closed Universes. Such discussion is connected with the physics of inflation, and its connection with vacuum decay. A unique realization of such ideas is baby-Universe formation by quantum tunneling which eventually disconnect from the parent spacetime \cite{Giddings:1988wv}. Recently, these ideas of connecting the physics of wormhole spacetimes to baby-Universes were revisited in \cite{Marolf:2020xie}, using features associated with a negative cosmological constant and asymptotically AdS boundaries.

Our main goal is to probe the ringdown of the aforementioned exact compact object solutions of scalar-tensor theory, by considering the propagation of linear test scalar perturbations minimally coupled to the metric, with the hope that these two objects are discernible, even though both possess a PS and effective AdS asymptotics. We will adopt the methodology used in \cite{Minamitsuji:2014hha,Dong:2017toi} and introduce a new minimally-coupled test scalar field in the gravitational action; the simplest case one can possibly envision. We note that the test scalar field we utilize is linear and should not be confused with the gravitational scalar of the theory, which backreacts to the metric to give rise to the scalarized BH and wormhole solutions we consider. By introducing a novel minimally-coupled linear test scalar field, we test the response of such solutions to small fluctuations, which in turn encode the information of the gravitational scalar that places a natural asymptotic effective boundary, although a negative cosmological constant is absent from the action. More complicated non-minimal couplings have been considered in such theories, though the effect of a coupling between the test and gravitational scalar leads to a critical coupling constant below which the QNM boundary conditions are not satisfied \cite{Dong:2017toi}.

The work is organized as follows. In Section \ref{sec:exact co solutions} we review the BH and wormhole solution with the non-minimal derivative coupling in the Horndeski scalar-tensor theory. In Section \ref{sec:Scalar-arbitrary-background} we derived the effective potentials for a test scalar field scattered off  from the BH and wormhole.  In Section \ref{sec:time domain profiles} we discuss the time-domain integration  scheme. In Section \ref{propagation} we study the propagation of the test scalar field in the background of the BH and the wormhole. Finally, in Section \ref{conclusion} we discuss our results and possible applications.
%---------------------------------------
%---------------------------------------
\section{Exact compact objects in scalar-tensor theory}\label{sec:exact co solutions}
%---------------------------------------
%---------------------------------------
In this section, we briefly review two exact compact object solutions \cite{Rinaldi:2012vy,Korolev:2014hwa} of the Horndenski Lagrangian with non-minimal kinetic coupling
\beq\label{action}
S=\int dx^4\sqrt{-g}\left\{\frac{\mathcal R}{8\pi} - \left[\varepsilon\,g_{\mu\nu}+ \eta\,
G_{\mu\nu}\right]\phi^{,\mu}\phi^{,\nu}\right\}~.
\eeq
Here, $\mathcal{R}$ is the Ricci scalar, $G_{\mu\nu}$ is the Einstein tensor, $g_{\mu\nu}$ is the metric tensor with $g=\det g_{\mu\nu}$, $\phi$ is a real massless scalar field and $\eta$ is a non-minimal coupling constant with dimensionality length squared.
In the case where $\varepsilon=1$, the theory contains a canonical scalar field with a positive kinetic term, while when $\varepsilon = -1$ the theory describes a phantom scalar field with a negative kinetic term.

%----------RINALDI BLACK HOLE-----------------
The BH solution found in \cite{Rinaldi:2012vy} is static, spherically symmetric and possesses an AdS-like boundary. The solution reads
\begin{align} \label{genmetric}
ds^2 &= -f(r)dt^2+g(r)dr^2+r^2(d\theta^2+\sin^2\theta d\varphi^2)~,
\end{align}
where
\begin{align}
 f(r) &= \frac{1}{4}\left( 3-\frac{8\mu}{r}+\frac{r^2}{3l_\eta^2}+\frac{l_\eta}{r}\arctan\frac{r}{l_\eta} \right) \,,\label{fbh}\\
 g(r) &= \frac{(r^2+2l_{\eta}^2)^2}{(r^2+l_{\eta}^2)^2 \,4 f(r)}\,,\label{gbh}\\[0.22cm]
 \Psi^2(r) &\equiv \left(\phi'(r)\right)^2 = -\frac{\varepsilon}{8\pi l^2_\eta}\,\frac{r^2(r^2+2l_{\eta}^2)^2}{(r^2+l_{\eta}^2)^3 \,4 f(r)}\,.\label{psibh}
\end{align}
 We would like to stress that this solution describes a BH only in the case where $\varepsilon=1$ and $\varepsilon\eta<0$ (see \cite{Rinaldi:2012vy,Korolev:2014hwa} for further details). Here, $\mu$ is an integration constant that plays the role of mass and $l_\eta=\sqrt{|\varepsilon\eta|}$ is a characteristic scale of the non-minimal coupling. The inverse tangent function restricts the domain of $r$ to $r\in (0,\infty)$. In the limit $r\rightarrow0$ the function $f(r)$ yields Schwarzschild asymptotics, i.e. $f(r)\approx 1-\frac{2\mu}{r}$, while for $r\rightarrow\infty$ one obtains $f(r)\approx\frac{3}{4}+\frac{r^2}{12 l^2_\eta}$ i.e. AdS-like asymptotics. Note that in the metric functions (\ref{fbh}) and (\ref{gbh}) the non-minimal coupling $l_\eta$ is present, which is the strength of the coupling of the scalar field to curvature. Therefore, the BH solution is dressed with a scalar field given in Eq. (\ref{psibh}).

%--------------SUSHKOV WORMHOLE----------------
The wormohole solution found in \cite{Korolev:2014hwa}, by following the approach of \cite{Rinaldi:2012vy}, reads\footnote{Note that the coordinates $(t, \xi, \theta , \phi)$ used here are not the Schwarzschild coordinates since, $\xi$ is not the curvature radius of a coordinate sphere $\xi=$const$>0$.}
\beq\label{whmetric}
ds^2=-f(\xi)dt^2+g(\xi)d\xi^2+(\xi^2+a^2)(d\theta^2+\sin^2\theta d\varphi^2)~.
\eeq
where $\varepsilon\eta<0$, $\varepsilon=-1$ and
\bea
g(\xi) &=& \frac{\xi^2(\xi^2+a^2+2l_\eta^2)^2}{(\xi^2+a^2)(\xi^2+a^2+l_\eta^2)^2 F(\xi)}~, \label{gwh}\\
f(\xi) &=& \frac{a}{\sqrt{\xi^2+a^2}}\exp\left[\int_0^\xi\frac{\xi(\xi^2+a^2+2l_\eta^2)^2} {l_\eta^2(\xi^2+a^2)(\xi^2+a^2+l_\eta^2)F(\xi)}d\xi\right]~, \label{fwh}\\
\Psi^2(\xi) &\equiv& \left(\phi'(r)\right)^2 = -\frac{\varepsilon}{8\pi l_\eta^2}
\frac{\xi^2(\xi^2+a^2+2l_\eta^2)^2}{(\xi^2+a^2)(\xi^2+a^2+l_\eta^2)^3 F(\xi)}~, \label{psiwh}
\eea
with
\beq
F(\xi)=3-\frac{8\mu}{\sqrt{\xi^2+a^2}}+\frac{\xi^2+a^2}{3l_\eta^2}
+\frac{l_\eta}{\sqrt{\xi^2+a^2}}\arctan \left(\frac{\sqrt{\xi^2+a^2}}{l_\eta}\right).
\eeq
Again, the non-minimal derivative coupling appears in the metric functions and   the phantom scalar field generating the wormhole is given in Eq. (\ref{psiwh}).
The function $F(\xi)$ has a minimum at $\xi=0$, thus to make it positive definite one should demand $F(0)>0$. Hence, one can derive the limitation on the upper value of the parameter mass parameter $\mu$
\beq\label{M}
2\mu<a\left(\frac34+\frac{\alpha^2}{12}+\frac{1}{4\alpha}\arctan\alpha\right)~,
\eeq
where $\alpha\equiv a/l_\eta$ is a dimensionless parameter which defines the ratio of the wormhole throat radius $a$ and the scale of the non-minimal kinetic coupling $l_\eta$. Far from the throat, in the limit $|\xi|\to\infty$, the metric functions $g(\xi)$ and $f(\xi)$ take the asymptotic form
\begin{equation}
g(\xi)=3\frac{l_\eta^2}{\xi^2}+O\left(\frac{1}{\xi^4}\right)~, \,\,\,\,\,\,\,\,\,\,\,\,\,\,\,\,
f(\xi)=A\frac{\xi^2}{l_\eta^2}+O(\xi^0)~,
\end{equation}
where $A$ depends on the parameters $a$, $l_\eta$, $\mu$ and can be calculated only numerically. These asymptotics correspond to AdS space with constant negative curvature. Close to the throat $\xi=0$ one finds
\begin{equation}
g(\xi)=B\frac{\xi^2}{l_\eta^2}+O(\xi^4)~, \,\,\,\,\,\,\,\,\,\,\,\,\,\,\,\,\,\,\, f(\xi)=1+O(\xi^2)~,
\end{equation}
where $B$ depends on $\alpha$ and $\mu$. Moreover, there is a coordinate singularity at $\xi=0$ where $g(0)=0$.

%------------------------------------------------
%------------------------------------------------
\section{Scalar field propagation on fixed gravitational backgrounds }\label{sec:Scalar-arbitrary-background}
%------------------------------------------------
%------------------------------------------------
In this study, we will be interested in the response of the compact object models described above against a linear test scalar field $\Phi$ minimally-coupled to the metric, but not the gravitational scalar $\phi$ in \eqref{action}. The dynamical propagation of a linear massless scalar perturbation $\Phi$ on the fixed background spacetime of a compact object, described by the metric tensor $g_{\mu\nu}$, is dominated by Klein-Gordon equation
\begin{equation}
\label{KleinGordon}
\square\Phi = 0 \Longleftrightarrow \frac{1}{\sqrt{-g}}\partial_\mu\left[\sqrt{-g} g^{\mu\nu} \partial_\nu \Phi \right]=0~.
\end{equation}
Due to spherical symmetry we can decompose the scalar field $\Phi(t,\rho,\theta,\phi)$ into a radial and angular parts, by introducing the ansatz
\begin{equation}
    \Phi(t,\rho,\theta,\phi) = \frac{\psi(\rho,t)}{R(\rho)}\,Y_{lm}(\theta,\phi)~, \label{scalar-ansatz}
\end{equation}
where $Y_{lm}$ are the standard spherical harmonics, $\rho$ is a general radial-like coordinate and $R(\rho)$\footnote{$R(\rho)\equiv R(r)=r$ for the BH and $R(\rho)\equiv R(\xi)=\sqrt{\xi^2+a^2}$ for the wormhole.} a function of $\rho$. Equation \eqref{KleinGordon} can, then, be recasted into a Schrodinger-like form
\begin{align}   \left[  \frac{\partial^2}{\partial t^2} - \frac{\partial^2}{\partial \rho_*^2} + V(\rho) \right] \psi(\rho,t) = 0~, \label{Schrod-generalform}
\end{align}
with the effective potential given by
\begin{align}
    V(\rho) = f(\rho)\left( \,\frac{\ell(\ell+1)}{R(\rho)^2} + \frac{R^{''}(\rho)}{g(\rho)R(\rho)} + \frac{f'(\rho) R'(\rho)}{2g(\rho)f(\rho)R(\rho)} - \frac{g'(\rho) R'(\rho)}{2 g^2(\rho) R(\rho)}\right)~, \label{Veff-generalform}
\end{align}
where $\ell$ is the angular quantum number and $\rho_\ast$ is the usual tortoise coordinate defined by
$$d\rho_* = \sqrt{\frac{g(\rho)}{f(\rho)}}\;d\rho\,.$$
Equation \eqref{Schrod-generalform} demonstrates that one is able to reduce the problem of scalar perturbations around compact objects into a single one-dimensional scattering problem with an effective potential. Applying this procedure on the BH \eqref{fbh}-\eqref{psibh} and the wormhole \eqref{gwh}-\eqref{psiwh} we find the corresponding effective potentials a test scalar field ``feels" when propagating on these backgrounds.

%---------------------------
\begin{figure}[t]
    \centering
    \includegraphics[scale=0.455]{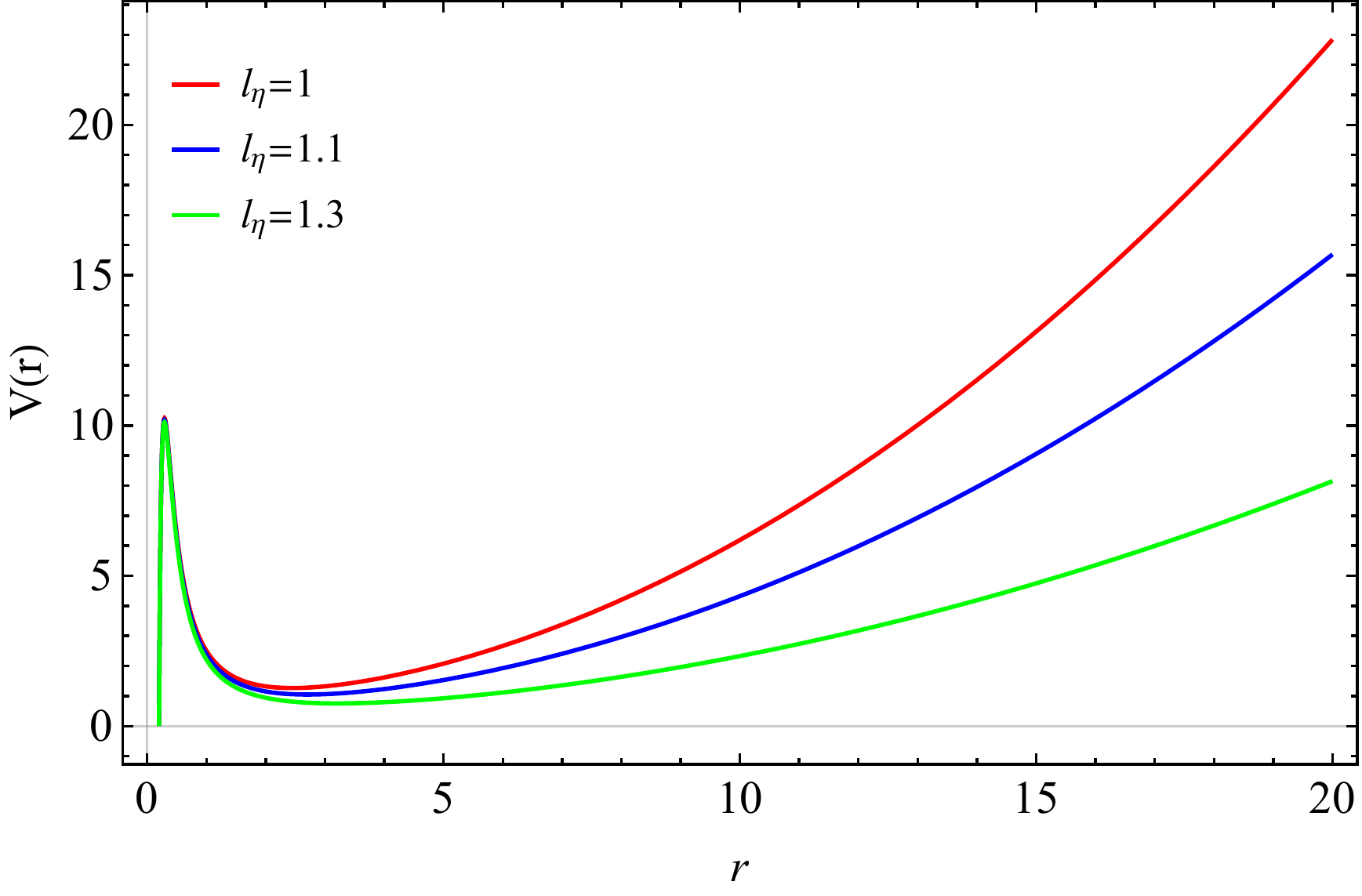}
    \includegraphics[scale=0.45]{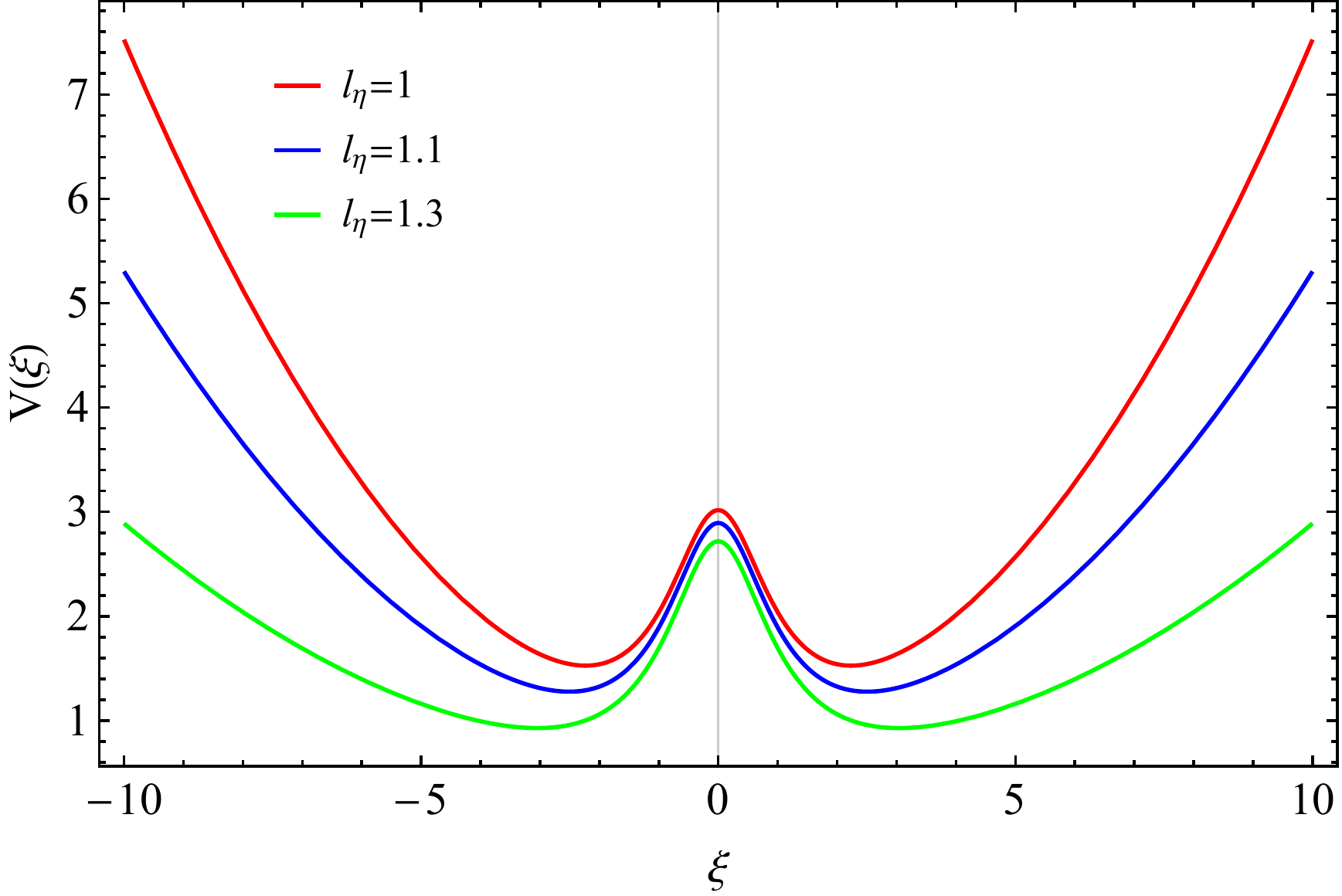}
    \caption{The effective potential of scalar perturbations with $\ell=1$ for the black hole \eqref{fbh}-\eqref{psibh} (left) and the wormhole \eqref{gwh}-\eqref{psiwh} (right) with throat radius $a=1$, for three different values of $l_\eta$ and $\mu=0.1$.} \label{fig:V-eff}
\end{figure}
%----------------------------
Fig. \ref{fig:V-eff} shows the effective potentials for various choices of non-minimal coupling constants. We observe that the BH spacetime has a peak right outside the event horizon, while asymptotically the effective potential diverges. Such asymptotic divergence encodes the AdS-like nature of the spacetime. The increment of $l_\eta$ leads to a more distant asymptotic boundary which could be explained from the fact that the non-minimal coupling has dimensionality length squared. In a sense, $l_\eta$ acts as an inverse cosmological constant, therefore at the limit $l_\eta\rightarrow\infty$ the spacetime becomes asymptotically flat, which would correspond to a zero cosmological constant.   The wormhole's effective potential is clearly different from that of the BH, as seen in Fig. \ref{fig:V-eff}. There is a single peak at the throat $\xi=0$, which corresponds to the PS, while asymptotically the potential diverges. The effect of $l_\eta$ is apparent in this case as well. Although not proven explicitly here, our numerics indicate that both peaks of $V$ occur close to the PS, since their amplitude is solely affected by $\ell$ which is directly associated with the energy of null particles trapped in unstable circular orbits at the PS.

%------------------------------------------------------------
\section{Time-domain integration scheme}\label{sec:time domain profiles}
%------------------------------------------------------------
In this section we briefly demonstrate the numerical scheme of time-domain integration, first proposed in \cite{Gundlach:1993tp}, which yields the temporal response of the scalar field as it propagates on a fixed background. By defining $\psi(\rho_\ast,t)=\psi(i\Delta \rho_\ast,j\Delta t)=\psi_{i,j}$, $V(\rho(\rho_*))=V(\rho_\ast,t)=V(i\Delta \rho_\ast,j\Delta t)=V_{i,j}$, equation \eqref{Schrod-generalform} takes the form
\begin{align}
    \frac{\psi_{i+1,j} - 2\psi_{i,j} + \psi_{i-1,j} }{\Delta\rho^2_\ast} - \frac{ \psi_{i,j+1} - 2\psi_{i,j} + \psi_{i,j-1} }{\Delta t^2} - V_i \psi_{i,j} = 0\,.
\end{align}
Then, by using as initial condition a Gaussian wave-packet of the form $\psi(\rho_\ast,t) = \exp\left[ -\frac{(\rho_\ast-c)^2}{2\sigma^2} \right]$ and $\psi(\rho_\ast,t<0) = 0$, where $c$ and $\sigma$ correspond to the median and width of the wave-packet, we can derive the time evolution of the scalar field $\psi$ by
\begin{align}
    \psi_{i,j+1} = -\psi_{i,j-1} + \left(\frac{\Delta t}{\Delta\rho_\ast}\right)^2\left( \psi_{i+1,j} + \psi_{i-1,j} \right) + \left( 2 - 2\left(\frac{\Delta t}{\Delta\rho_\ast}\right)^2 - V_i \Delta t^2 \right) \psi_{i,j}
    \label{psi-evolution}\,,
\end{align}
where the Von Neumann stability condition requires that $\frac{\Delta t}{\Delta \rho_\ast} < 1$.
Moreover, the effective potential is positive and vanishes at the event horizon (but not at the wormhole throat), however, it diverges as $r\to\infty$ (or $|\xi|\to\infty$). This requires that $\psi$ should vanish at infinity for both compact objects in study, which corresponds to reflective boundary conditions.  To calculate the precise values of the potential $V_i$, we integrate numerically the equation for the tortoise coordinate and then solve with respect to the corresponding radial coordinate. Various convergence tests were performed throughout our numerical evolution, with different integration steps and precision, to reassure the validity of our ringdown profiles.

%----------------------------------
\section{Propagation of perturbations on compact objects in scalar-tensor theory} \label{propagation}
By applying the numerical procedure outlined above, we calculate the temporal response of linear massless scalar field perturbations on the BH and wormhole solutions discussed. In what follows, we assume the mass of both compact objects to $\mu=0.1$ (if not stated otherwise) and obtain the perturbation response at a position arbitrarily close to the event horizon for the BH, and at $\xi=0.01$ for the wormhole.

\subsection{Black hole}
%----------------------------------
\begin{figure}[H]
    \includegraphics[scale=0.445]{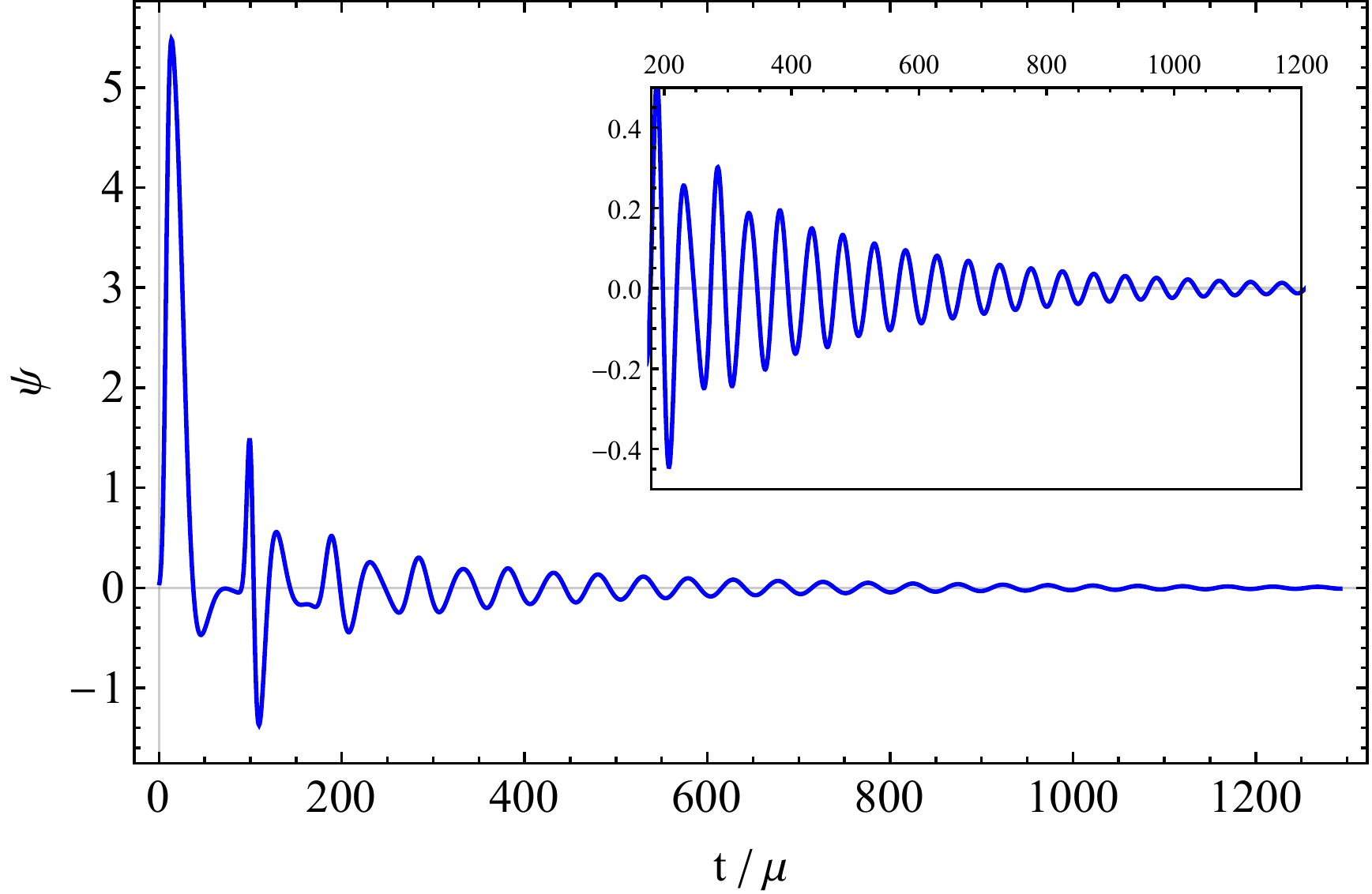}\hskip 2ex
    \includegraphics[scale=0.478]{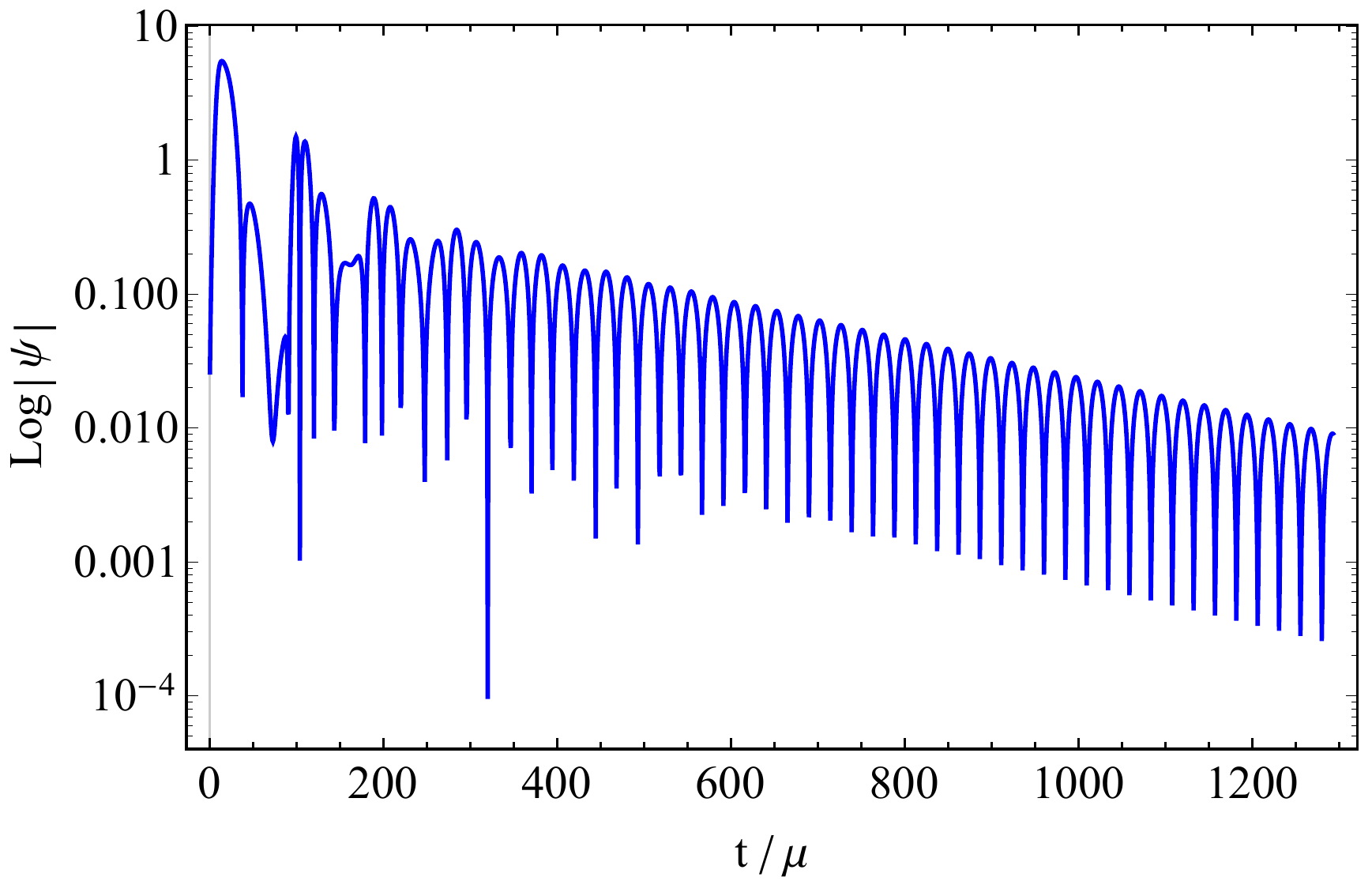}\vskip 1ex
    \includegraphics[scale=0.445]{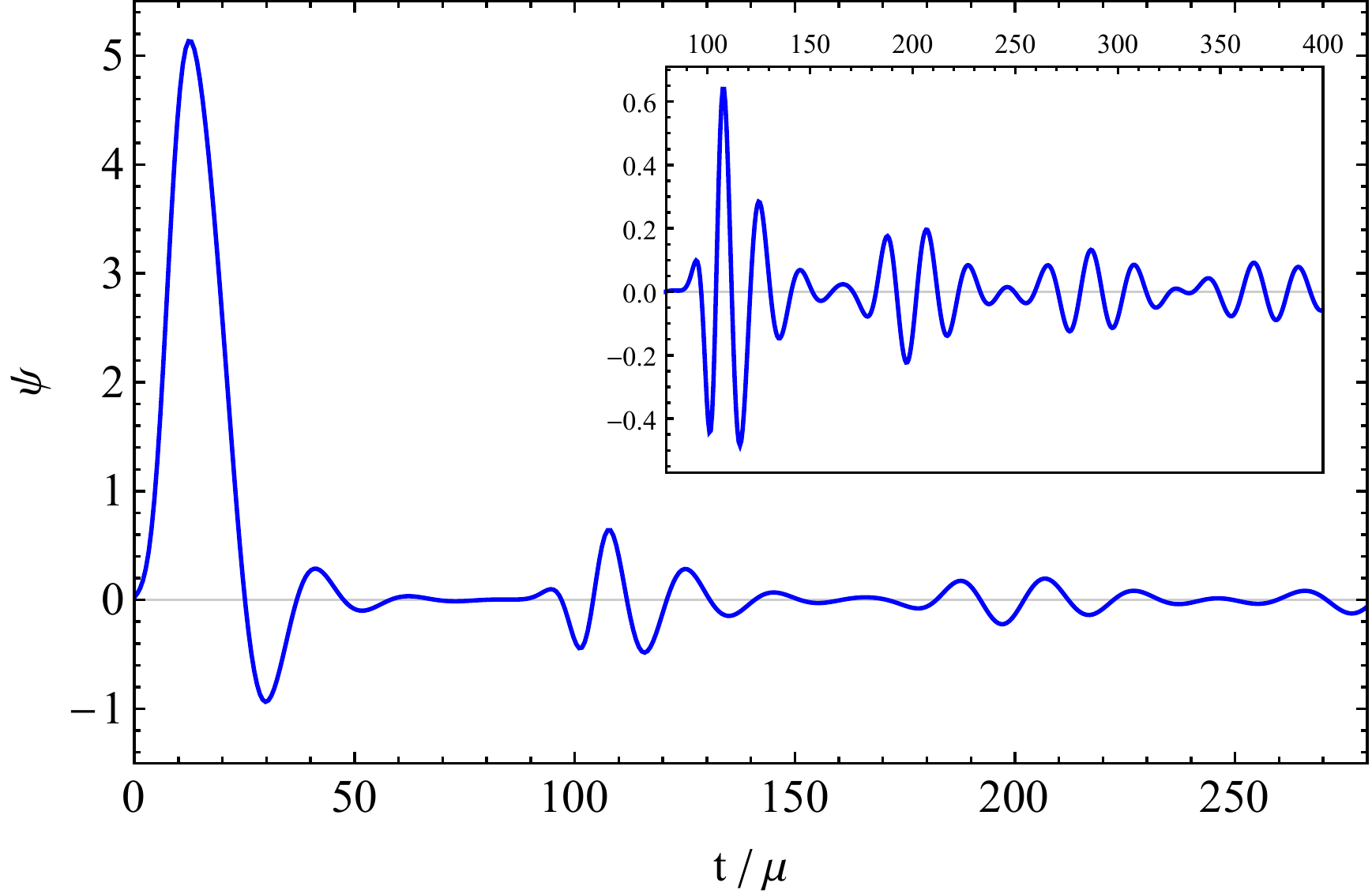}\hskip 2ex
    \includegraphics[scale=0.478]{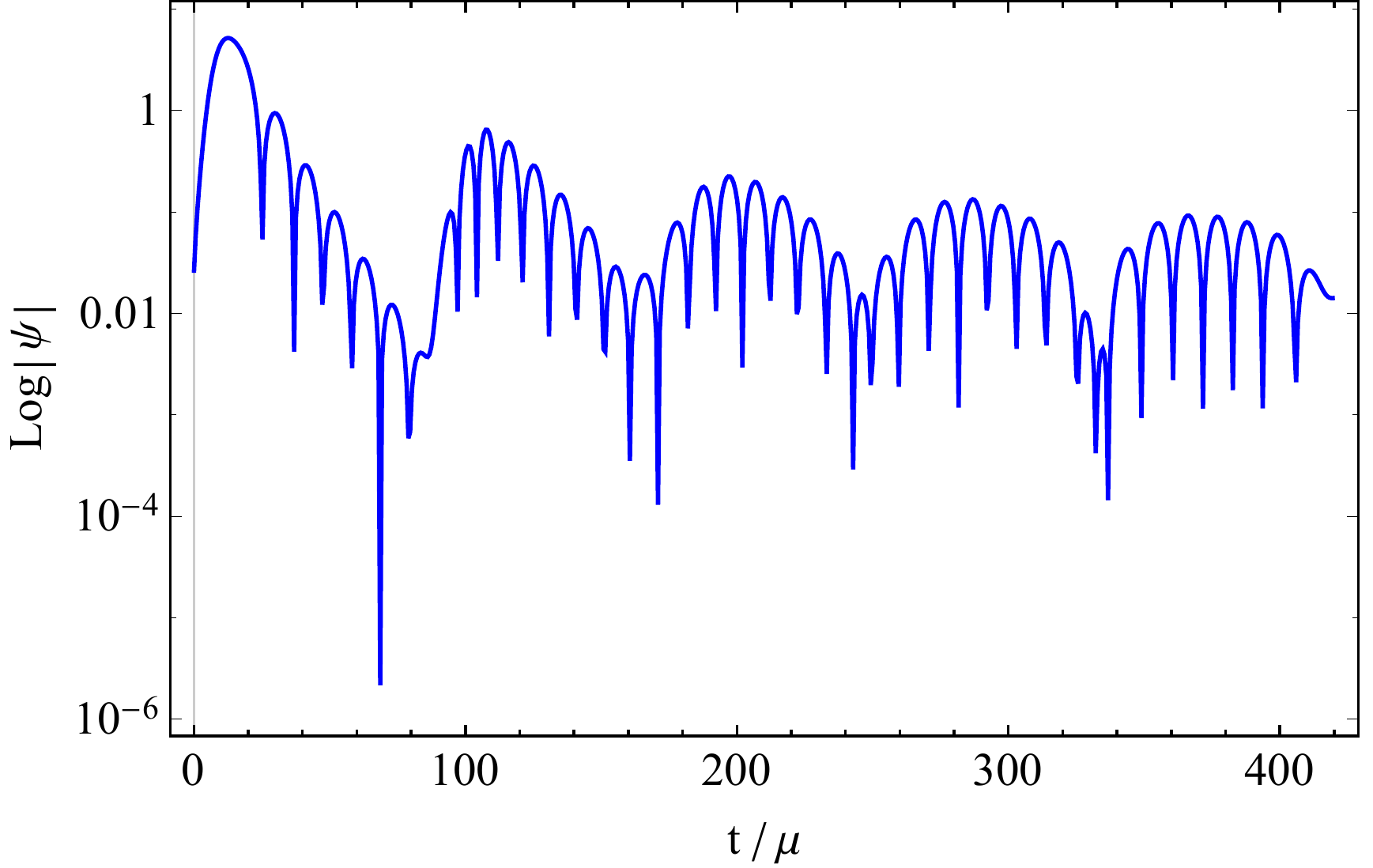}\vskip 1ex
    \includegraphics[scale=0.445]{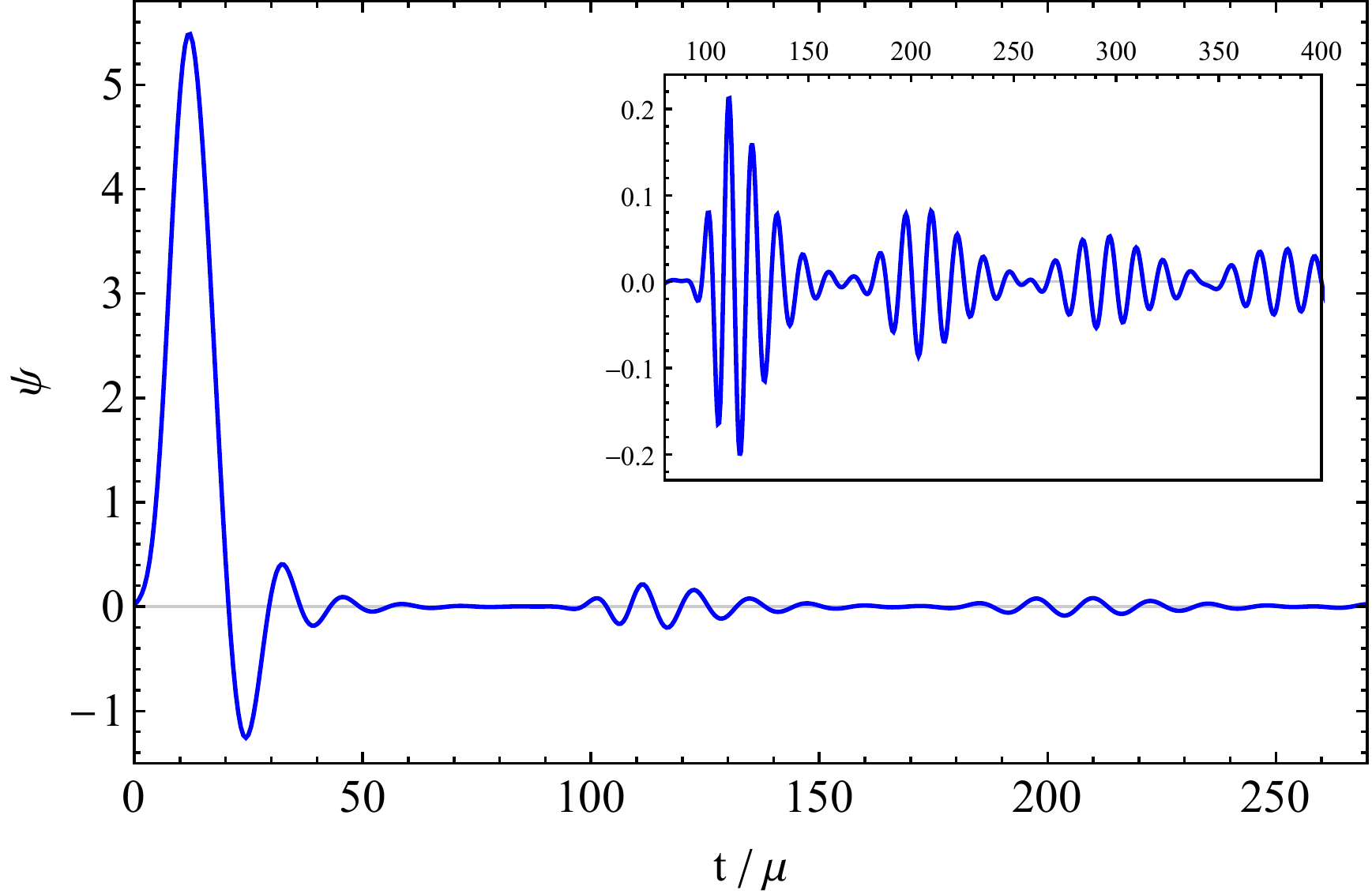}\hskip 2ex
    \includegraphics[scale=0.478]{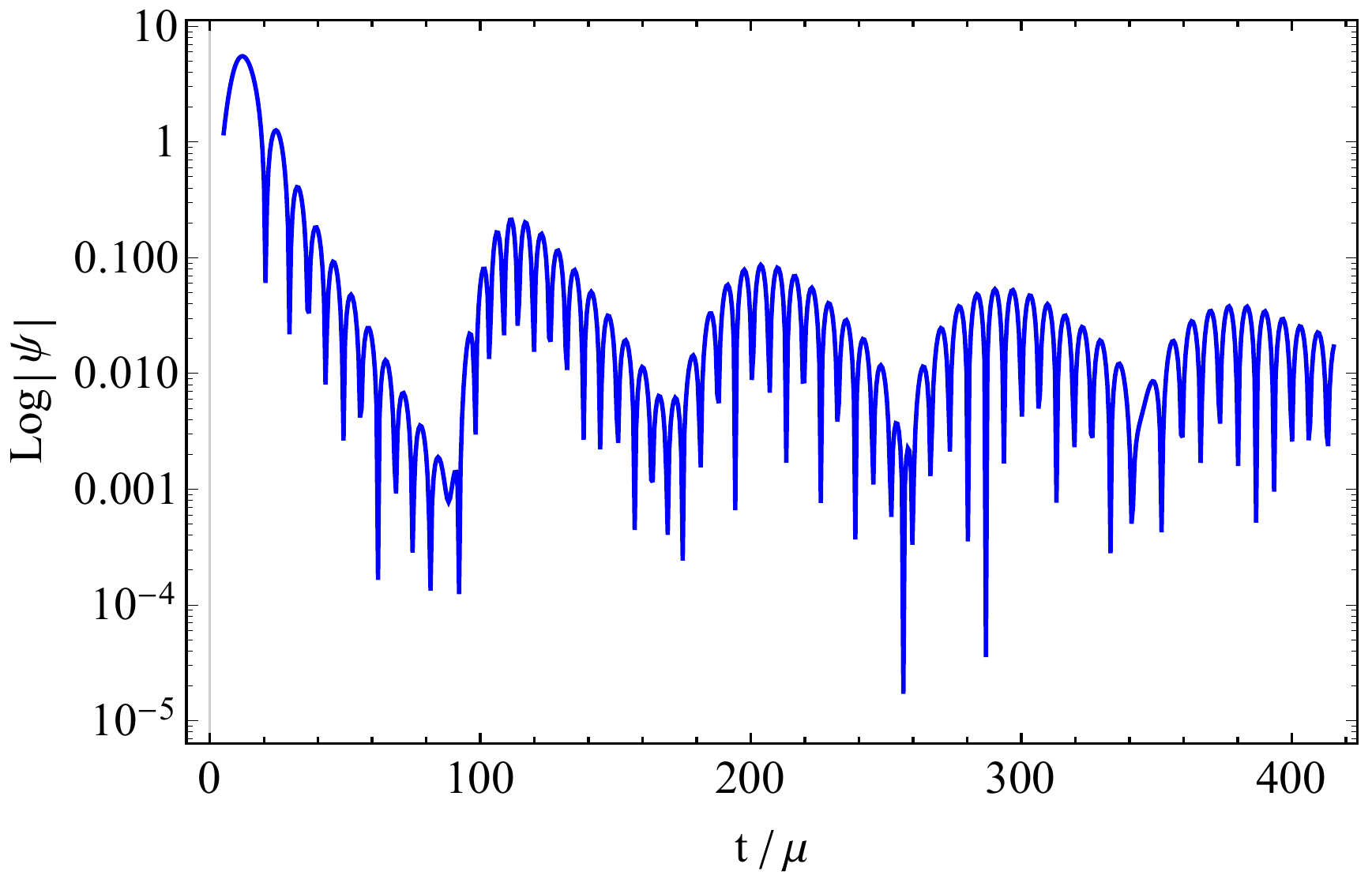}
    \caption{Time evolution of scalar perturbations with $\ell=0$ (top panel) $\ell=1$ (middle panel) and $\ell=2$ (bottom panel) on the black hole background \eqref{fbh}-\eqref{psibh} with $\mu=0.1$ and $l_\eta=1$.}
    \label{fig:TDP-Rin-l012}
\end{figure}

Fig. \ref{fig:TDP-Rin-l012} displays the evolution of a linear scalar perturbation field on the background of the BH solution \eqref{fbh}-\eqref{psibh}. The most obvious effect we can observe is the emergence of echoes following the initial quasinormal ringdown. This pattern becomes more evident for higher $\ell$ due to the fact that more energy is carried away from the PS when perturbed. For spherically-symmetric $\ell=0$ perturbations, on the other hand, the echo pattern is not so evident since the field does not excite the PS significantly, therefore the echoes fall off rapidly. Our investigation confirms that the decay rate of scalar perturbations follows an exponential fall-off, as in \cite{Horowitz:1999jd,Wang:2000dt}, which is more evident for the case of $\ell=0$. This behaviour is in contrast to that of asymptotically flat BH perturbations, where the quasinormal ringing gives way to a power-law cutoff \cite{Gundlach:1993tp,Leaver:1986gd,Gundlach:1993tn}, and its occurrence is related to the asymptotic nature of timelike infinity in AdS spacetimes which serves as a reflective boundary. The echoes have significantly smaller amplitudes compared with the initial ringdown, which is in agreement with the studies in \cite{Minamitsuji:2014hha,Dong:2017toi,Abdalla:2018ggo} and the dissipative nature of the event horizon, designating modal stability. 

It is worthy to note that further analytical investigations of perturbations in AdS BHs led to the conclusion that solutions of the Klein-Gordon equation with fixed angular quantum number $\ell$, indeed decay exponentially \cite{Holzegel:2011uu}. However,  an accumulation of all solutions, possessing finite energy, achieves a logarithmic decay rate, due to the presence of stable trapping \cite{Holzegel:2013kna} (for a discussion of the wave equation in the interior of AdS BHs see \cite{Kehle:2018zws,Kehle:2020zfg}).

\begin{figure}[t]
    \centering
    \includegraphics[scale=0.47]{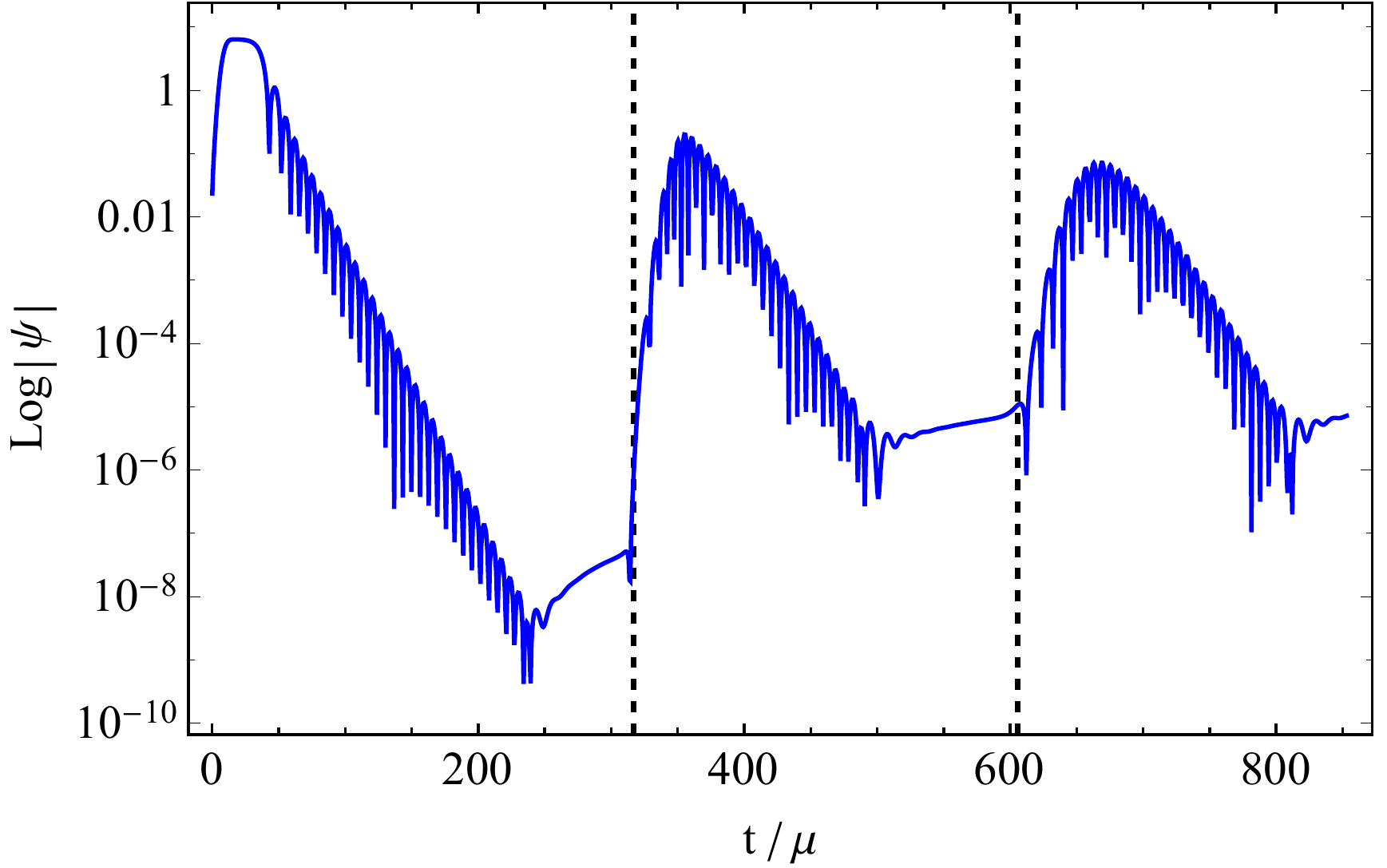}
    \includegraphics[scale=0.47]{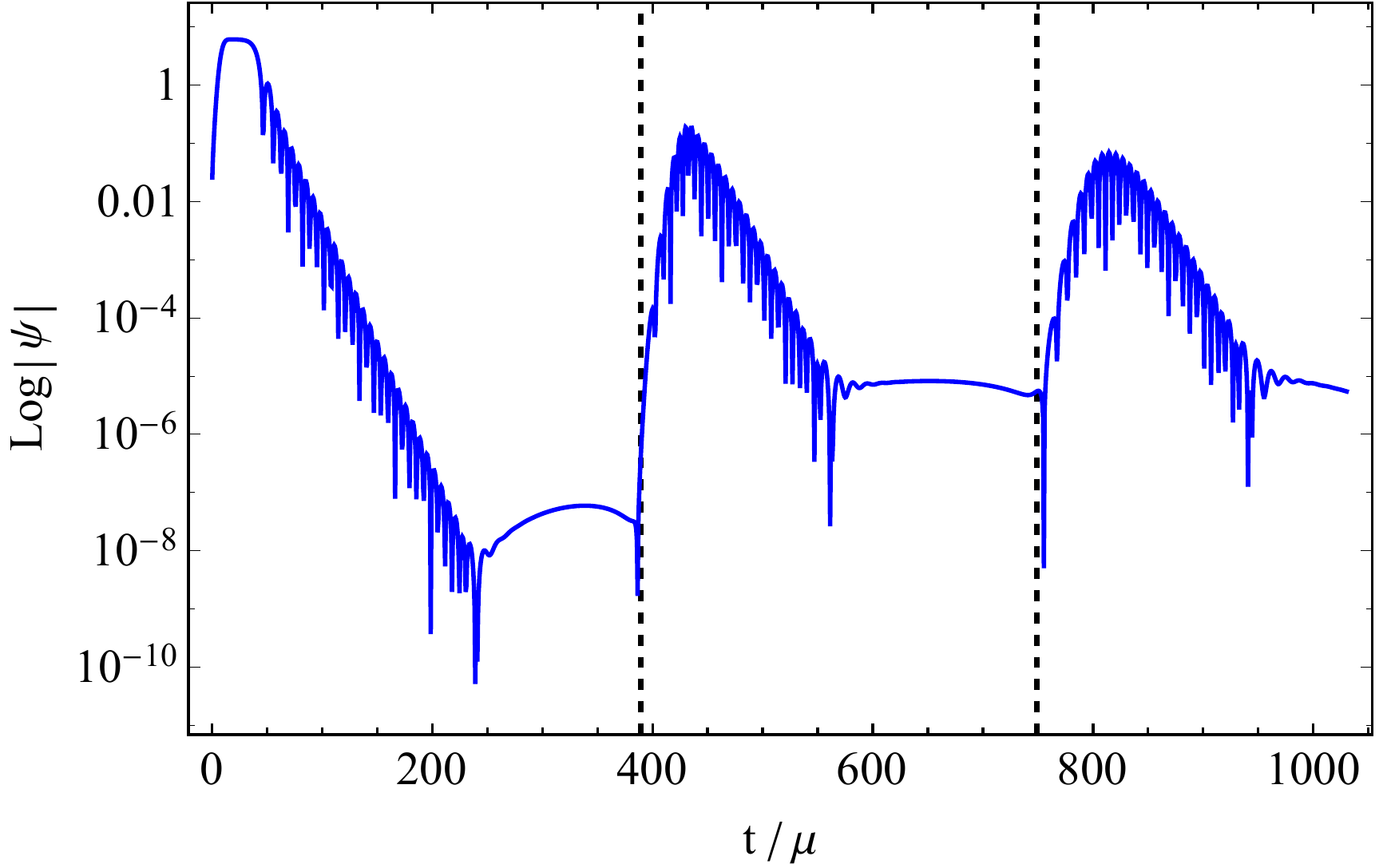}
    \caption{Time evolution of scalar perturbations with $\ell=2$ on the black hole background \eqref{fbh}-\eqref{psibh} with $\mu=0.1$ and $l_\eta=4$ (left), $l_\eta=5$ (right). The black vertical dashed lines indicate the time one expects the next echo to arrive, as calculated from \eqref{echoe-time}.}
    \label{fig:TDP-Rin-leta3,4,10-l2}
\end{figure}

In Fig. \ref{fig:TDP-Rin-leta3,4,10-l2} the effect of the derivative coupling constant is illustrated. When $l_\eta$ increases the effective AdS boundary moves further away from the event horizon (see Fig. \ref{fig:V-eff}). As a consequence, the scalar wave reflected off the PS has to travel a larger distance before it reaches the reflective AdS boundary and return to re-perturb the PS. Thus, the increment of $l_\eta$ leads to a delay of the echoes. It is important to note that the echoes appear in this case, not due to trapping of waves between the PS and the surface of the compact object, but rather due to the asymptotic nature of infinity. This effect may have important implications in AdS/CFT correspondence, if an actual negative cosmological constant is included, where a ringdown in the bulk corresponds to the approach to thermal equilibrium in the boundary CFT though a sequence of ringdown signals, such as echoes, does not yet have a proper boundary interpretation (though see \cite{Dey:2020lhq,Dey:2020wzm}).

To justify our statement we have computed numerically the time interval needed for light to perform a round trip from the PS to the AdS boundary. For a metric of the form \eqref{genmetric} the characteristic time-scale is given by \cite{Cardoso:2016rao,Saraswat:2019npa}
\begin{align}
    \Delta t = 2 \int_{PS}^{Boundary} \sqrt{\frac{g(r)}{f(r)}}\;dr\label{echoe-time}\,.
\end{align}
As can be seen from Fig. \ref{fig:TDP-Rin-leta3,4,10-l2}, the temporal location of the echoes, as obtained from the numerical integration, is in good agreement with the values of $\Delta t$ calculated from Eq. \eqref{echoe-time} (shown with dashed lines in Fig. \ref{fig:TDP-Rin-leta3,4,10-l2}). This agreement further supports that the formation of echoes is due on the secondary perturbations of the PS from the reflected scalar field on the effective AdS boundary.

%-------------------------------
\subsection{Wormhole}
\begin{figure}[H]
    \includegraphics[scale=0.468]{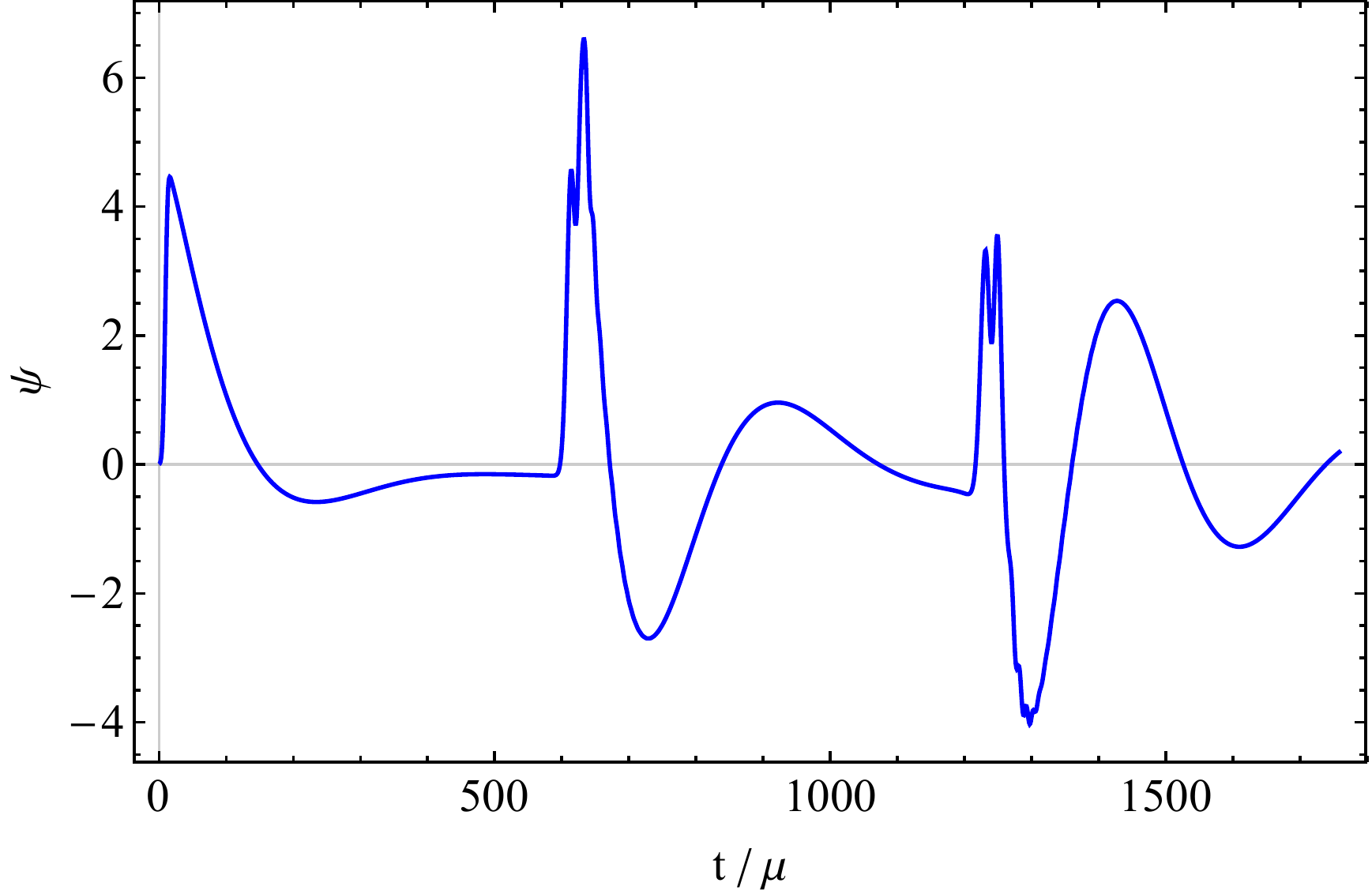}\hskip 2ex
    \includegraphics[scale=0.478]{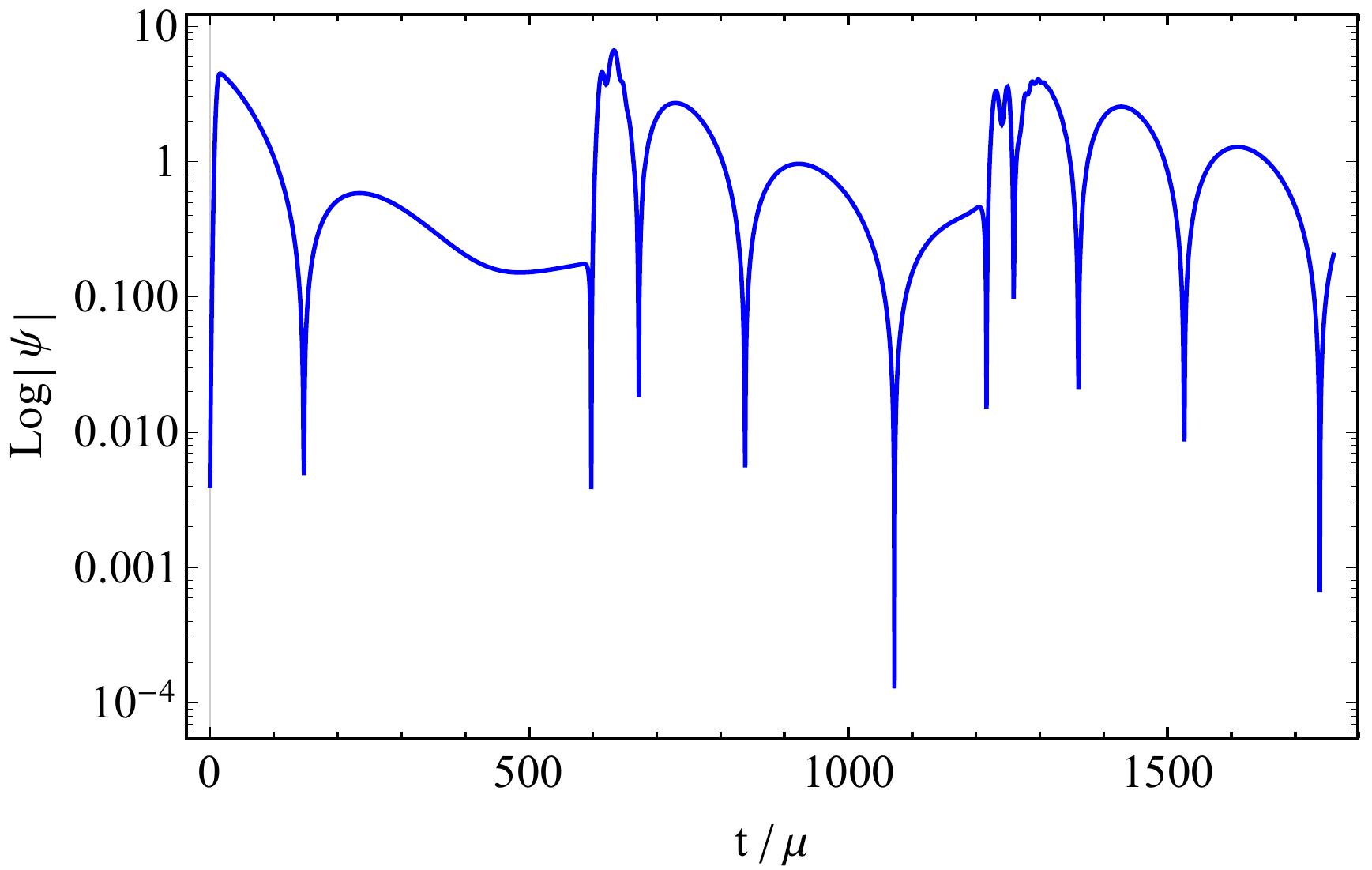}\vskip 1ex
    \includegraphics[scale=0.468]{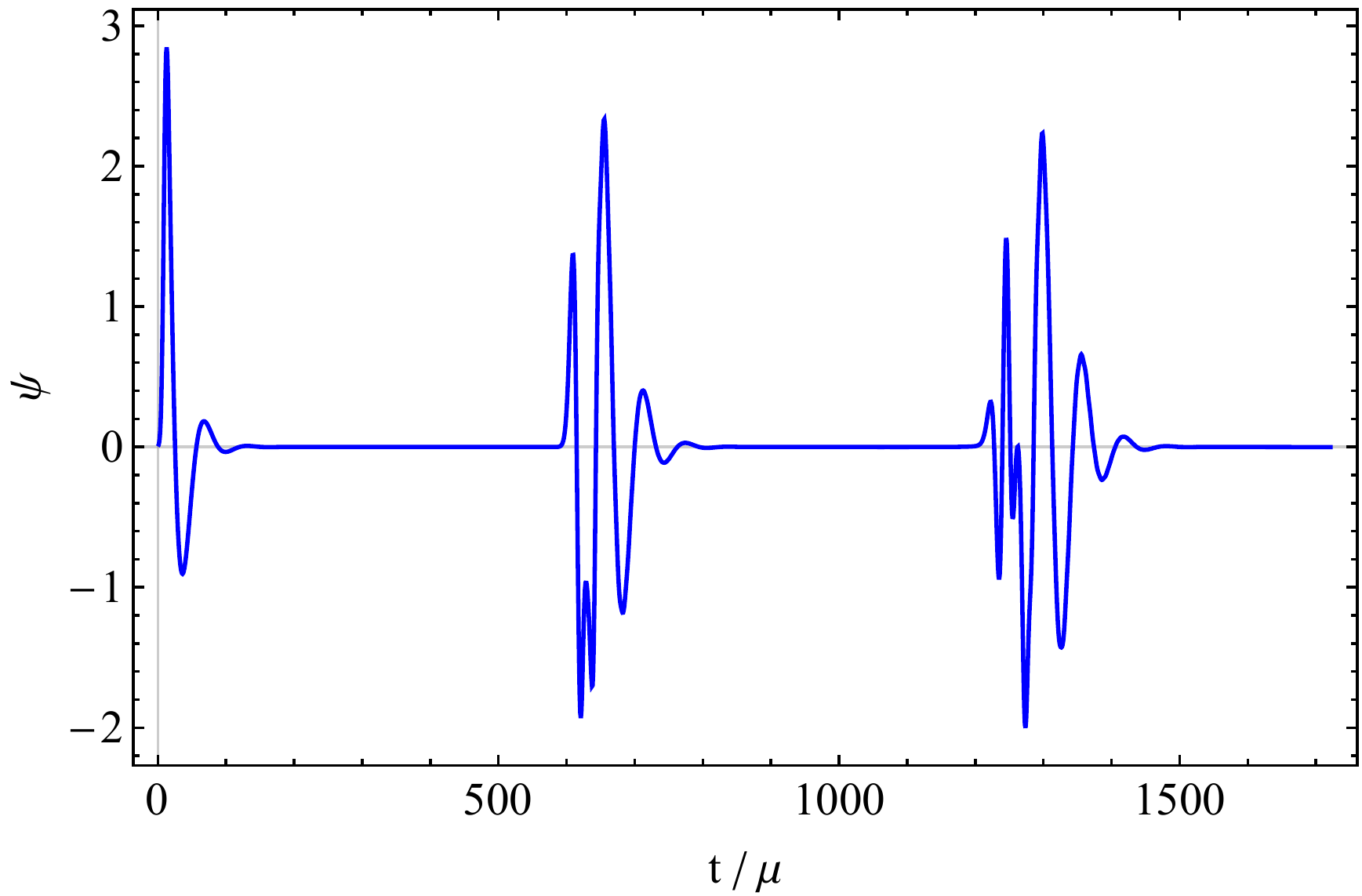}\hskip 2ex
    \includegraphics[scale=0.478]{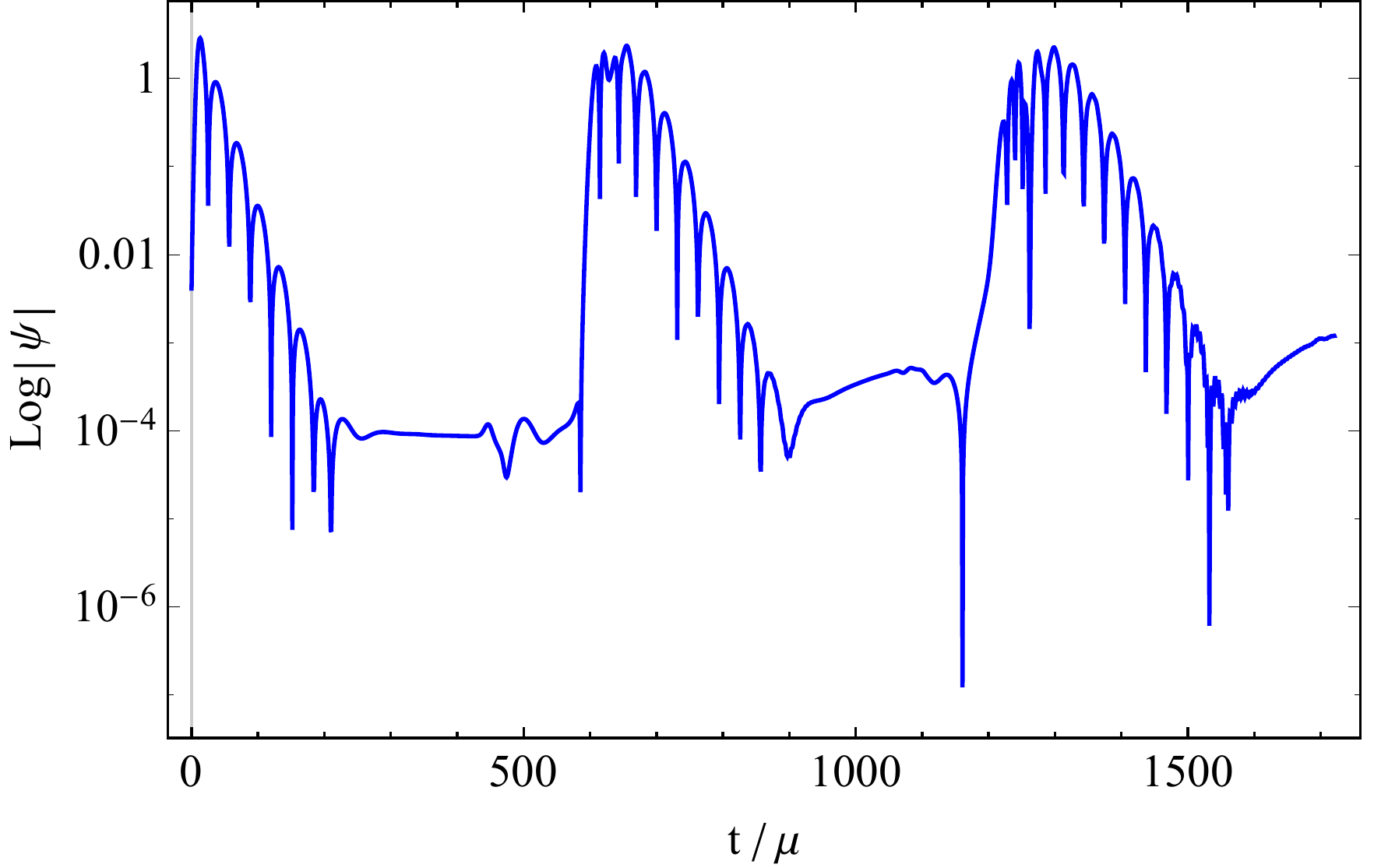}\vskip 1ex
    \includegraphics[scale=0.468]{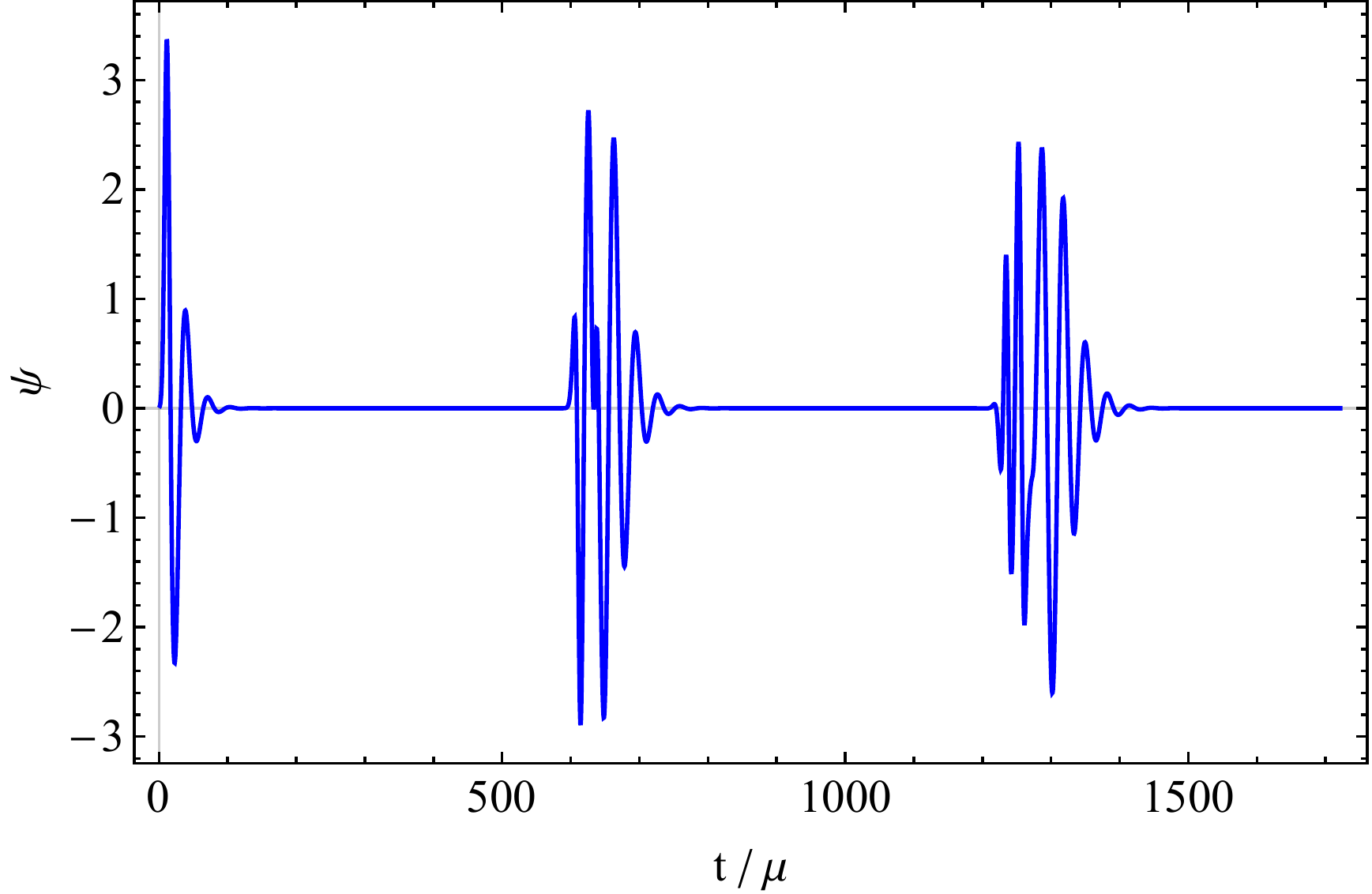}\hskip 2ex
    \includegraphics[scale=0.478]{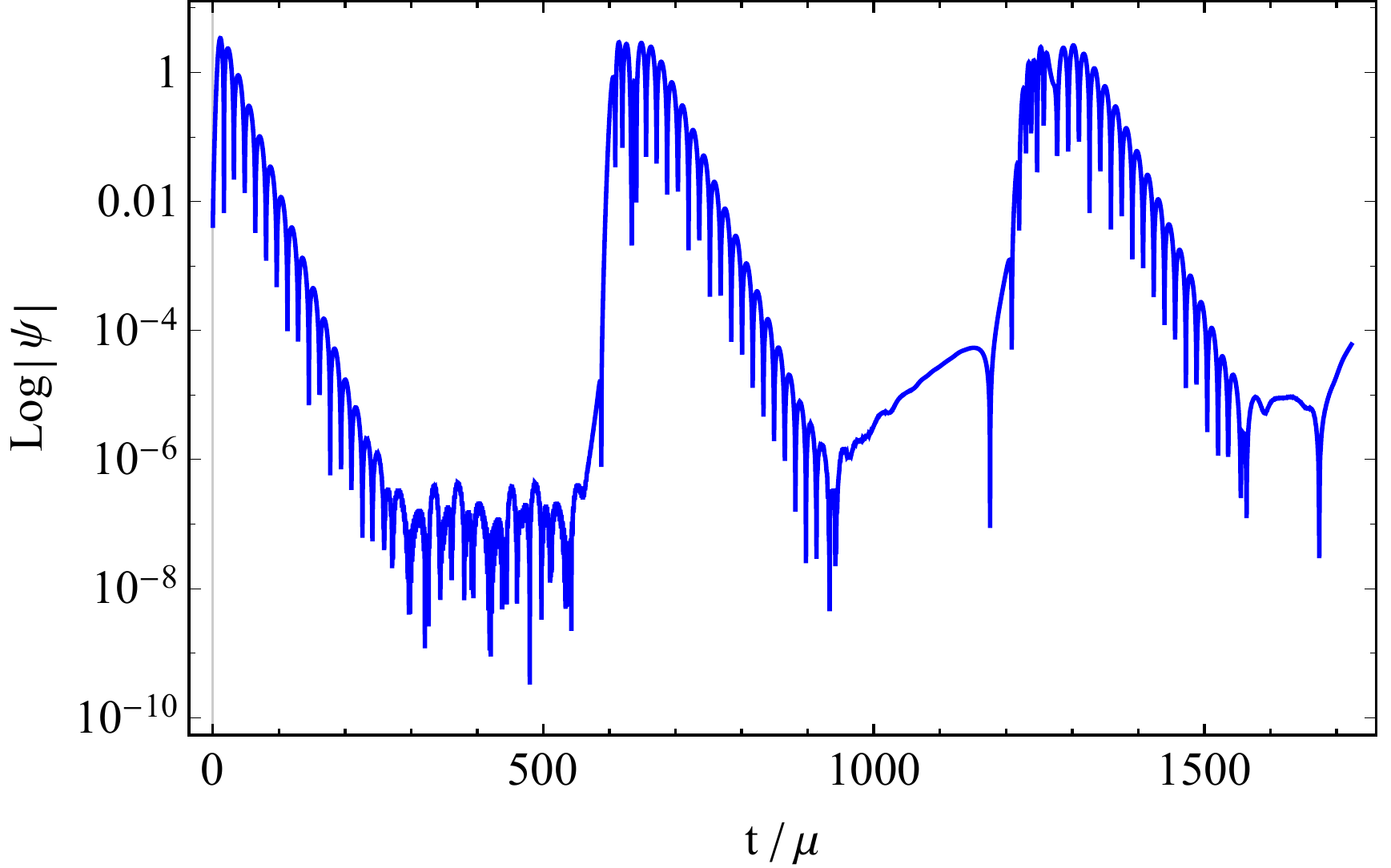}
    \caption{Time evolution of scalar perturbations with $\ell=0$ (top panel) $\ell=1$ (middle panel), $\ell=2$ (bottom panel) on the wormhole background \eqref{fwh}-\eqref{psiwh} with $l_\eta=10,\,\mu=0.1$ and $a=1$.}
    \label{fig:TDP-Sus-leta10-l012}
\end{figure}

In Fig. \ref{fig:TDP-Sus-leta10-l012} we demonstrate the behavior of the test scalar field as it propagates in the wormhole solution \eqref{fwh}-\eqref{psiwh}. The temporal response exhibits echoes, as in the BH case above, which follow the initial ringdown due to the first encounter of the probe field with the PS. In a similar manner, the $\ell=0$ perturbations do not significantly excite the PS of the wormhole, thus the echoes are not as oscillatory as the ones obtained for $\ell>0$. However, we notice that the amplitude of the echoes does not decrease with time, in contrast to the response in the BH setup.

Through this effect, one then can easily distinguish if the compact object is a BH or wormhole. The underlying mechanism that leads to such a behavior could be understood from the fact that instead of an event horizon we have a wormhole throat, therefore, energy cannot be dissipated. The probe field travels through the throat and into the second Universe, to be reflected back from the second effective AdS boundary (see Fig. \ref{fig:V-eff}). The small ``glitches" shown in echoes of Fig. \ref{fig:TDP-Sus-leta10-l012} (in the linear scale) appear due to the fact that we measure the response of the test field at $\xi=0.01$, thus the reflected wave from the AdS boundary of the primary Universe arrives slightly earlier than the reflected wave from the boundary of the secondary Universe, to superpose and lead to an echo of equal amplitude to that of the initial outburst.

The effect of the non-minimal coupling $l_\eta$ is shown in Fig. \ref{fig:TDP-Sus-leta2.5,5,10}. Besides the fact that the initial ringdown and the echoes have the same amplitude, one can realize that $l_\eta$ serves as a scale of the Universe, since for higher $l_\eta$ values, the field has to travel a larger distance from the throat to the AdS boundary and back. This results to a proportionality between the coupling $l_\eta$ and the echo time. To illustrate this, we have approximated the time interval between two echoes using the relation
\begin{align}
	\label{echo_time_worm}
	\Delta t =  2 \int_{Throat}^{Boundary} \sqrt{\frac{g(r)}{f(r)}}\;dr~.
\end{align}
The agreement between the resulting signal from time integration and the characteristic time from Eq. \eqref{echo_time_worm} (see vertical dashed lines in Fig. \ref{fig:TDP-Sus-leta2.5,5,10}) justifies further that the echoes are produced due to the presence of the effective AdS boundary and not due to the existence of a double barrier effective potential that usually appears in wormhole solutions (see \cite{Cardoso:2016rao}, \cite{Bronnikov:2019sbx}-\cite{Liu:2020qia}).

%-------------------------------
% ----------------------------

The existence of echoes of equal amplitude to that of the initial ringdown is an indication that such compact objects may possess normal modes of oscillation, similar to those found in \cite{Evnin:2015gma,Fierro:2018rna,Anabalon:2019lzc}. In fact, one could perform a mode decomposition on the test scalar field to calculate these modes, though the complicated form of the metric components render such analysis rather challenging. Since the echoes seem to possess a characteristic timescale \eqref{echo_time_worm}, it is intriguing to approximate the normal modes as $\omega\sim 2\pi/\Delta_t$. Our numerics indicate that $\omega\sim\mu/l_\eta$, with $\ell$ not playing a dominant role in this approximation, besides affecting the oscillation rate of each ringdown. A modal analysis would shed more light to the validity of our approximation and to the existence of normal modes in such wormholes.

\begin{figure}[t]
	\centering
	\includegraphics[scale=0.475]{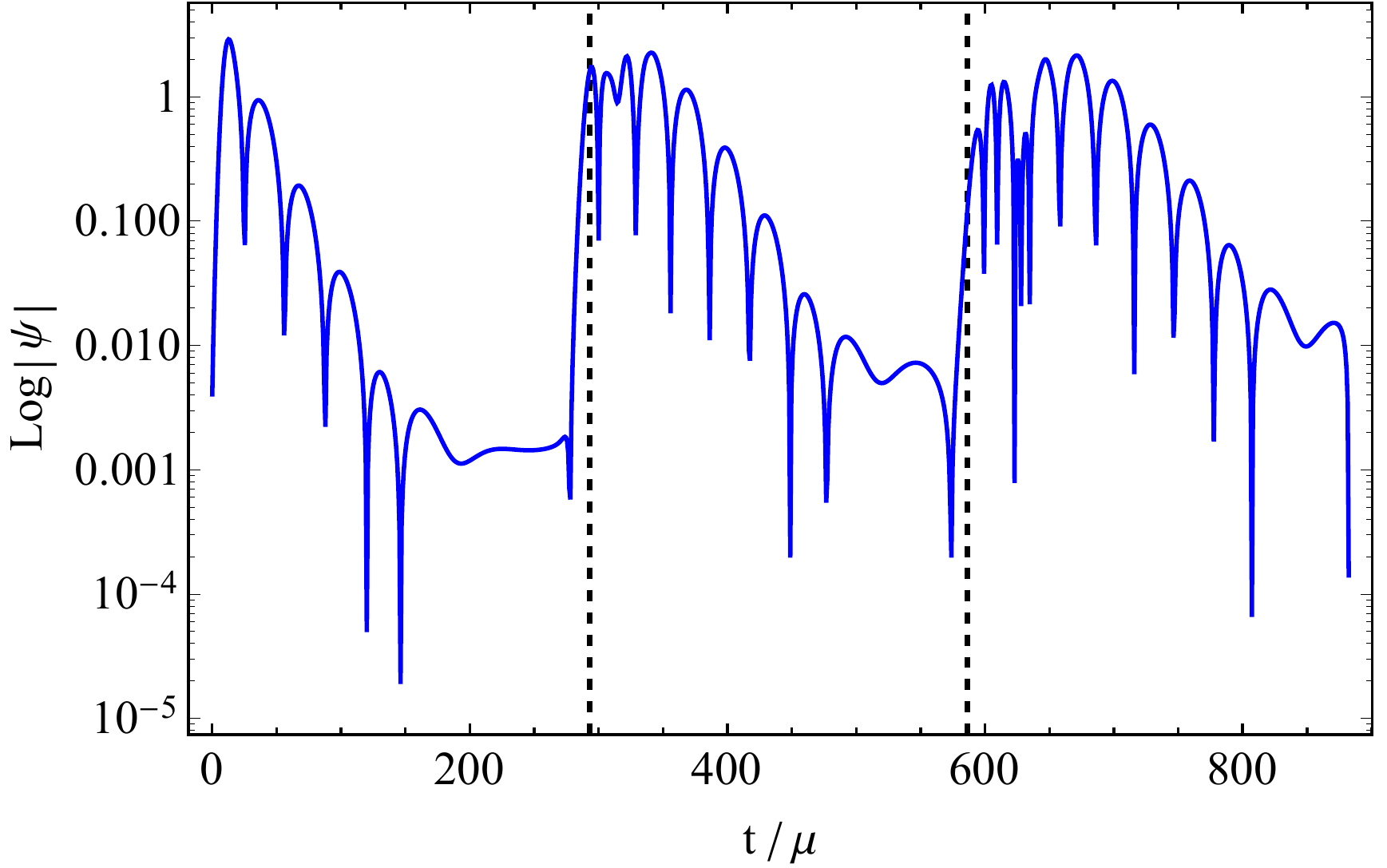}\hskip 1ex
	\includegraphics[scale=0.47]{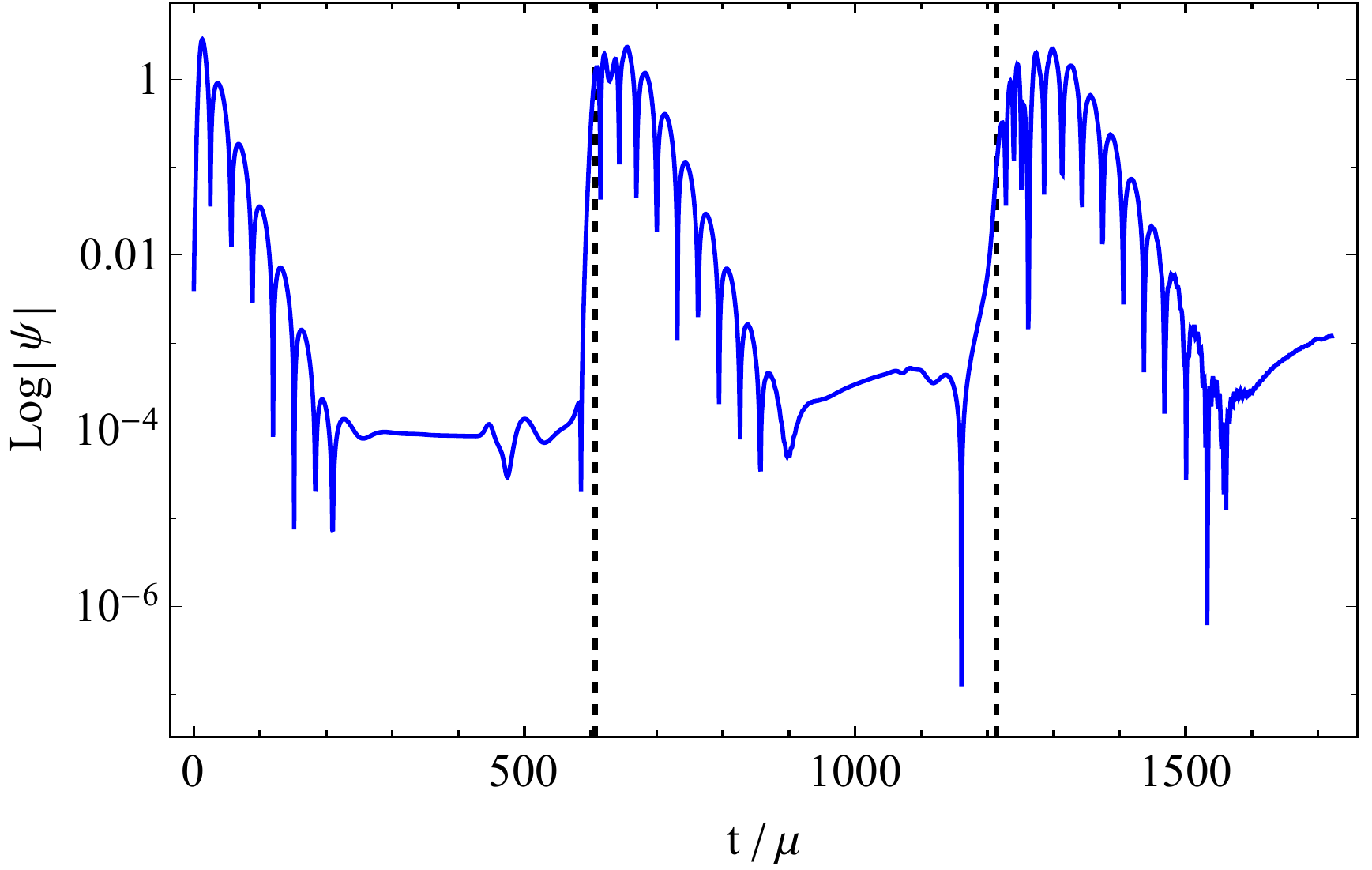}
	\caption{Time evolution of scalar perturbations with $\ell=1$ on the wormhole background \eqref{fwh}-\eqref{psiwh} with $\mu =0.1$, $a=1$ and
		$l_\eta=5$ (left), $l_\eta=10$ (right).}
	\label{fig:TDP-Sus-leta2.5,5,10}
\end{figure}

\subsubsection{Extremal wormhole}

\begin{figure}[t]
     \centering
     \includegraphics[scale=0.47]{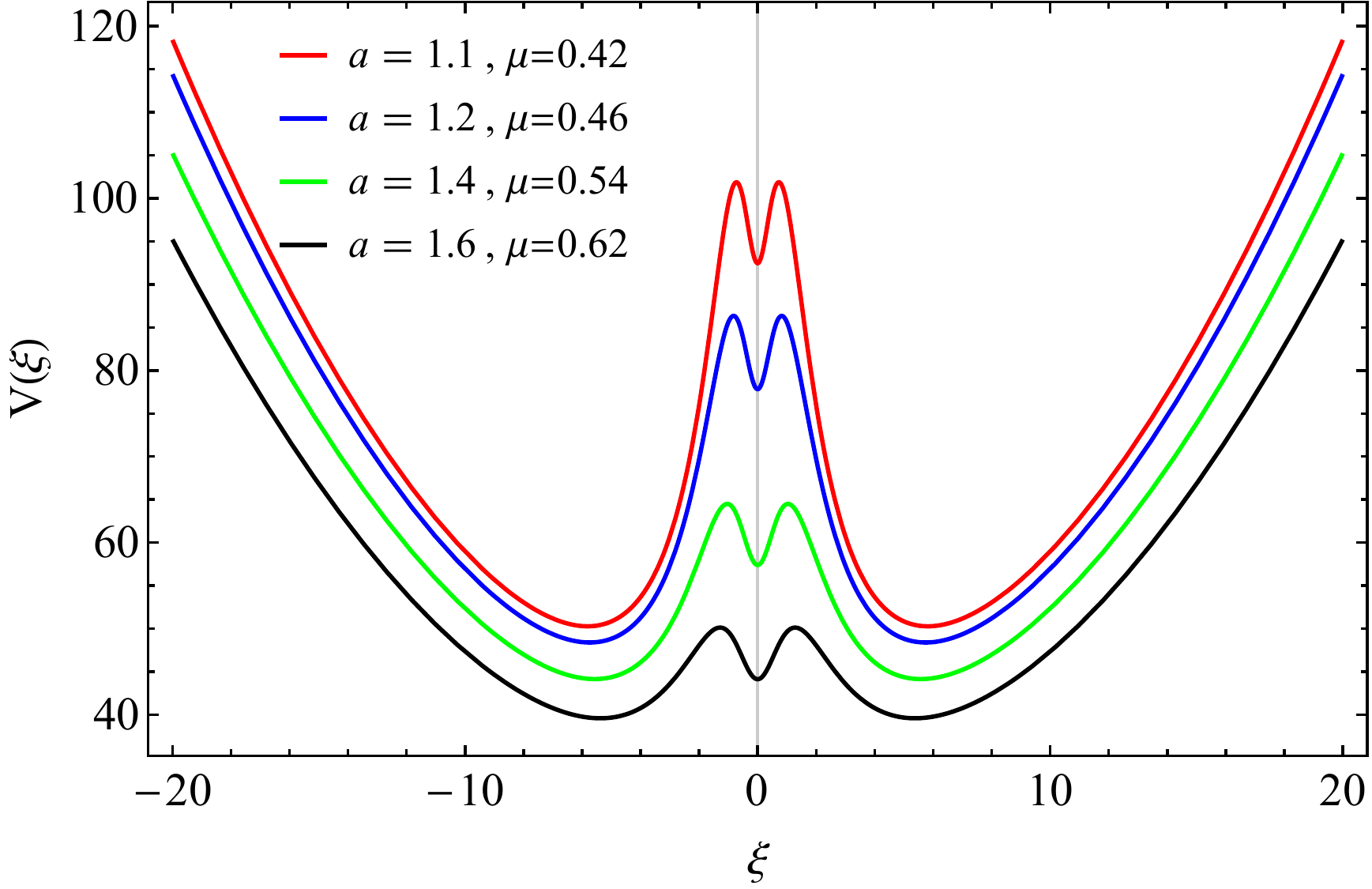}\hskip 1ex
     \includegraphics[scale=0.47]{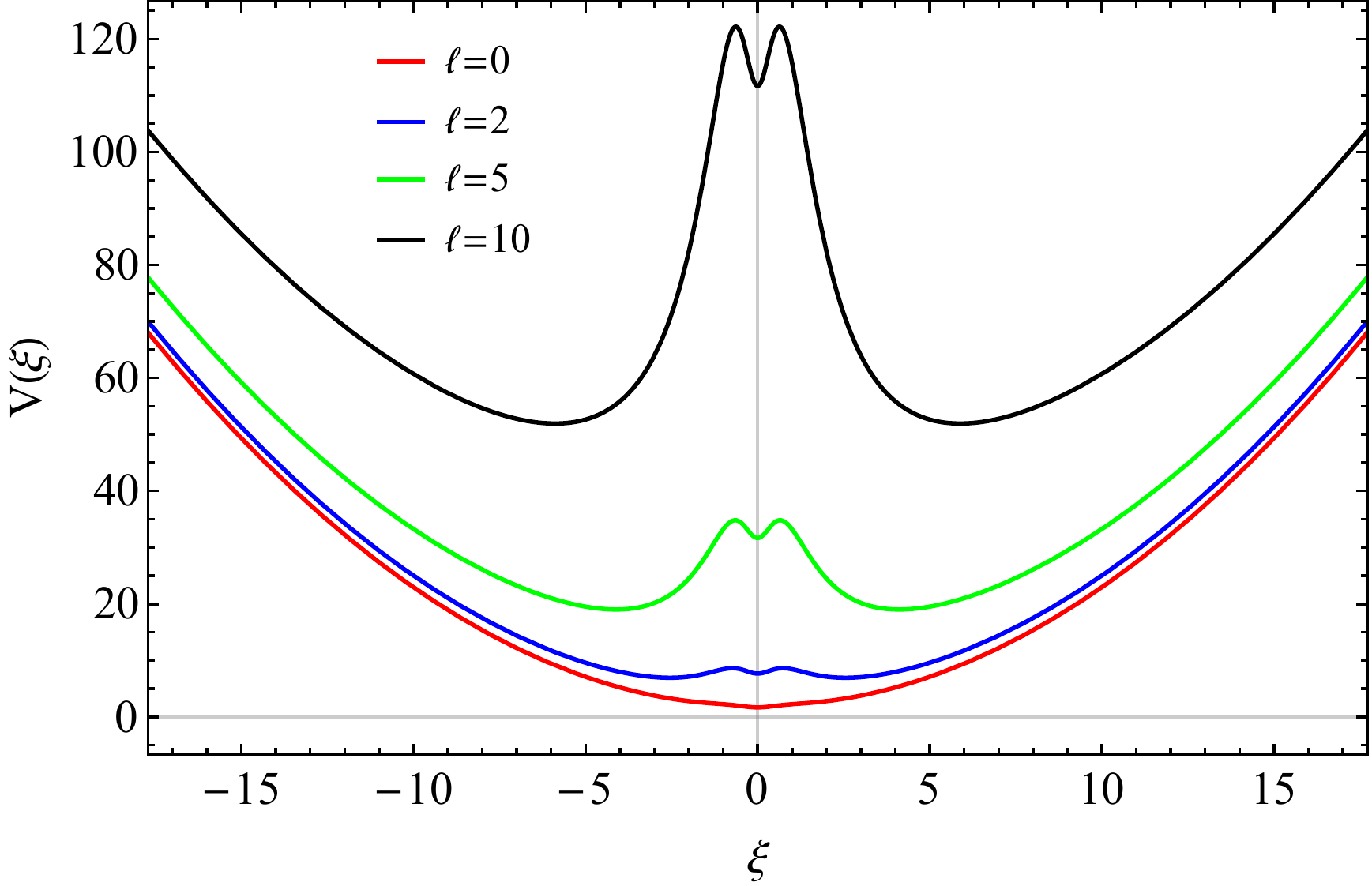}
     \caption{Left panel: Effective potential with $l_\eta=1\,,\,\ell=10$ and $\mu=\mu_{extreme}$ for each value of the throat radius $a$. Right panel: Effective potential with $l_\eta=1\,,\,a=1\,,\,\mu=\mu_{extreme}$ for various angular momenta $\ell$.}
     \label{extremal pot}
\end{figure}

So far, the wormhole parameters considered give rise to a single peak at the effective potential, which leads to two trapping regions separated by a potential barrier. If we consider a different set of parameters, where the wormhole mass is nearly extremal $\mu\simeq\mu_{extreme}$ (or exactly extremal), then the potential peak splits into two potential barriers, for large angular momenta. In this case, three trapping regions appear, which depend on the throat radius, near-extremal mass an angular momentum in a manner which is demonstrated in Fig. \ref{extremal pot}. The existence of three potential wells in the wormhole background will lead to echoes which arise from two different regions: the primary region between the PS and the asymptotic AdS boundary, and the novel secondary region between the wormhole throat.

 \begin{figure}[t]
     \centering   \includegraphics[scale=0.46]{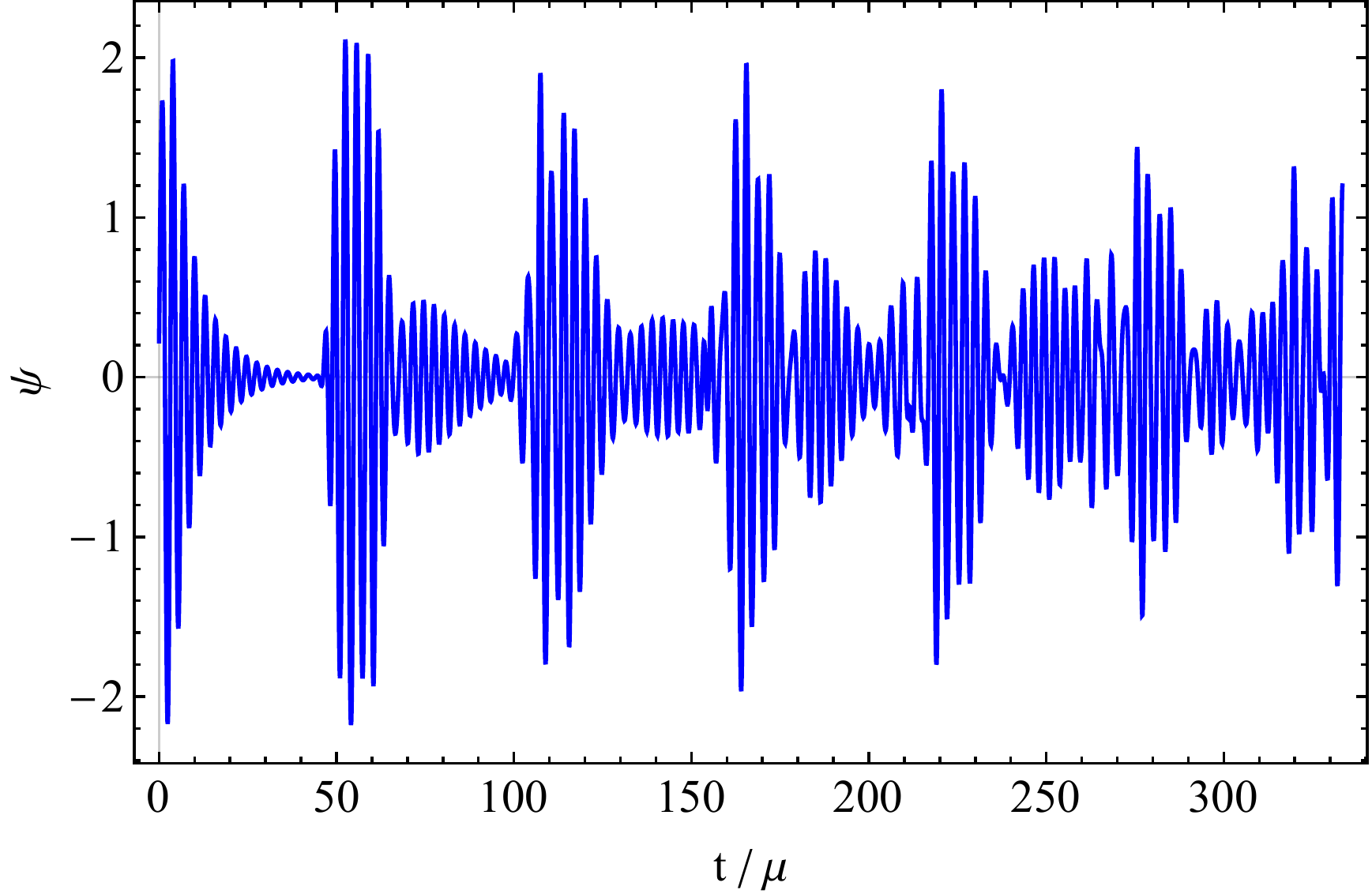}\hskip 1ex
     \includegraphics[scale=0.48]{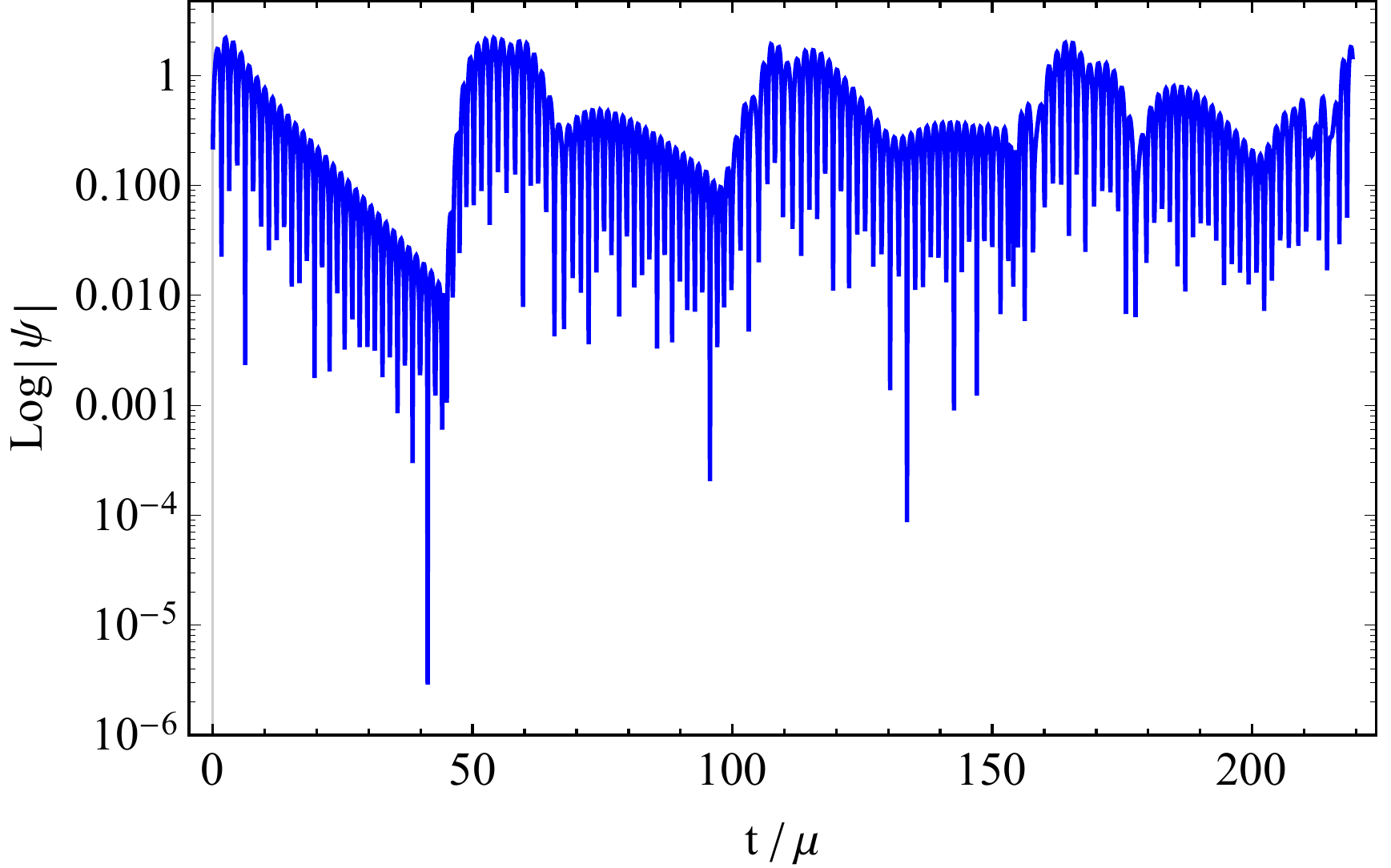}\vskip 1ex
     \includegraphics[scale=0.47]{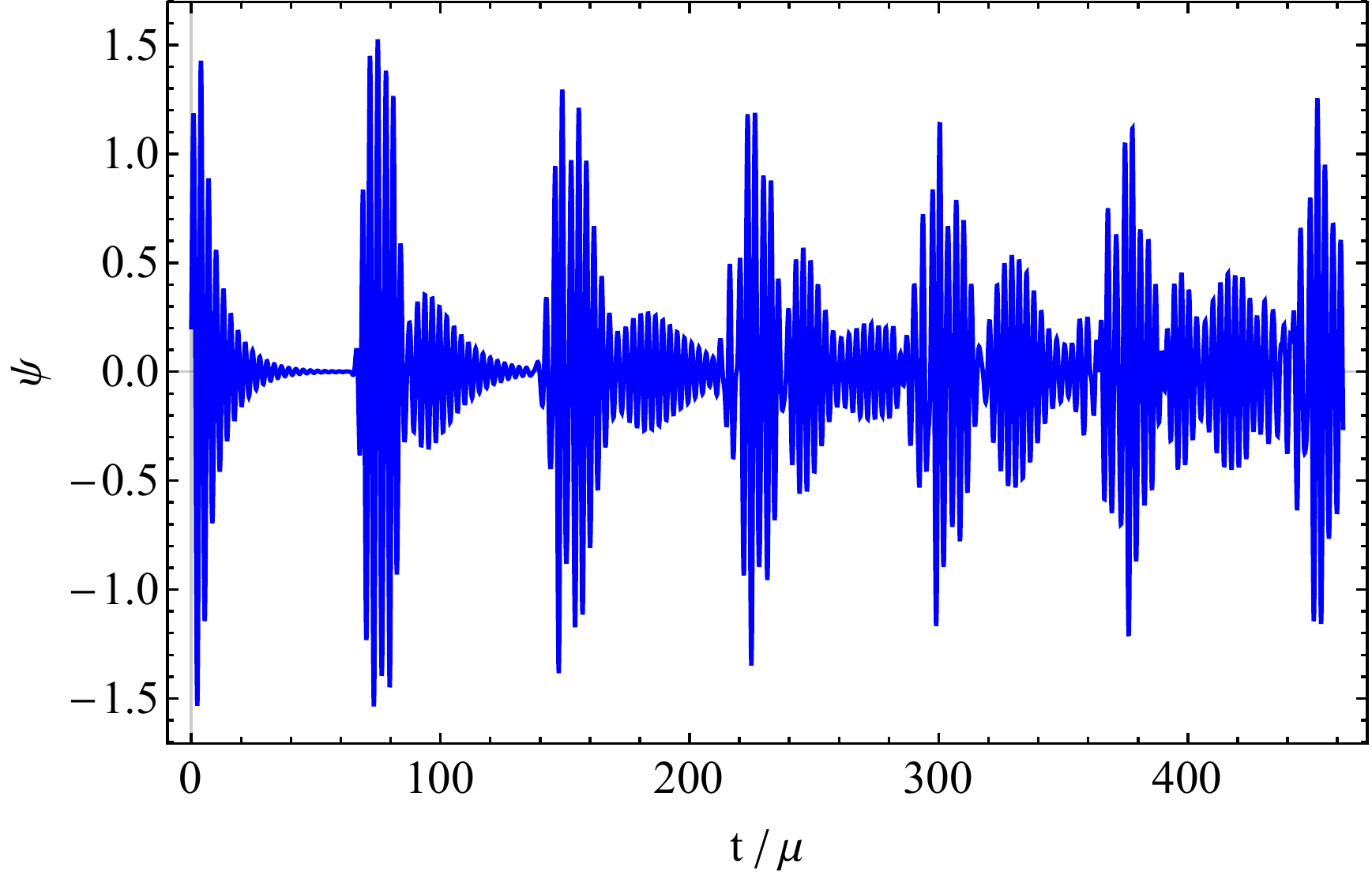}\hskip 1ex
     \includegraphics[scale=0.48]{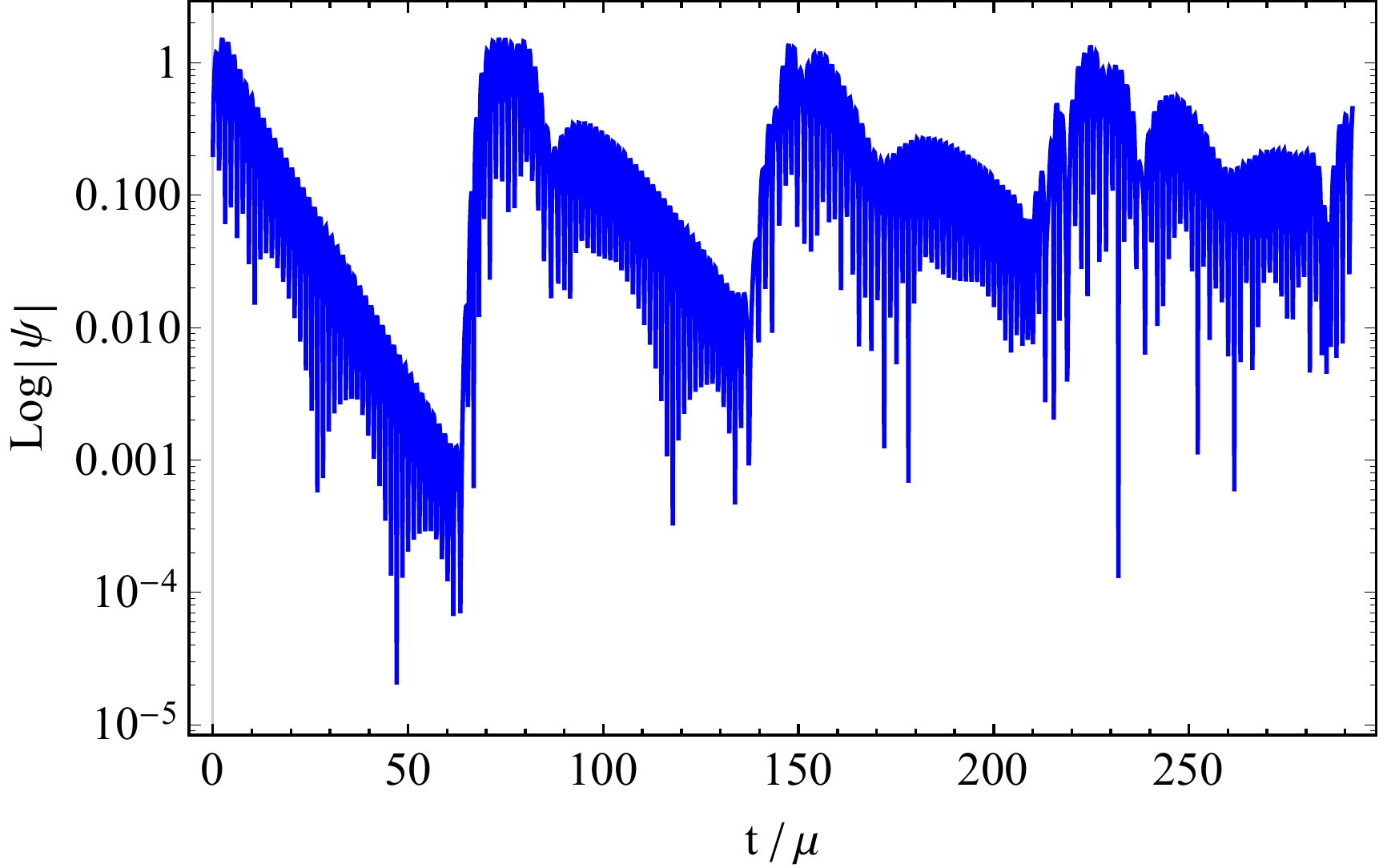}
     \caption{Time evolution of scalar perturbations with $\ell=5$ on the wormhole background with $\mu=\mu_{extreme}\approx0.37$, $a=1$ and $l_\eta=5$ (top panel), $l_\eta=7$ (bottom panel).}
     \label{extremal time}
 \end{figure}

 In Fig. \ref{extremal time} we observe a new qualitative behaviour of linear scalar perturbations which is not present in the non-extremal wormhole setups considered above. Besides the primary echoes arising from the reflection of the test field at the AdS boundary, a new series of echoes appears in between them, with smaller amplitude than the primary ones. These secondary echoes are a product of the new trapping region at the wormhole throat. Although the first couple of secondary echoes are quite visible in Fig. \ref{extremal time}, at sufficiently late times the superposition of primary and secondary echoes renders them hardly distinguishable. Nevertheless, the existence of primary and secondary echoes, associated with different trapping regions, is reported here for the first time in a wormhole spacetime with trapping regions that arise naturally.

\section{Conclusions}
\label{conclusion}

The GW ringdown, where the final object relaxes to a stable state, contains key information about the perturbed compact object's externally observably quantities. Recently, it has been argued \cite{Cardoso:2016rao,Cardoso:2016oxy} that the late-time ringdown signal may incorporate signatures for the existence of ECOs, such as wormholes, or horizon-scale quantum corrections \cite{Foit:2016uxn,Cardoso:2019apo,Agullo:2020hxe} (though see \cite{Coates:2019bun}), in the form of echoes.

Here, we considered minimally-coupled test scalar perturbations on exact BH and wormhole solutions of scalar-tensor theory, which possess an effective negative cosmological constant, leading to AdS asymptotics, due to the presence of a non-minimally coupled `gravitational' scalar to the Einstein tensor. We find that similar effects arise in the late-time ringdown waveform for both compact objects under study. After the initial ringdown, the test field response exhibits echoes, with timescales proportional to the non-minimal coupling constant. Although the BH considered here does not contain any quantum corrections at the event horizon, the effective asymptotic AdS boundary, that the gravitational scalar introduces to the scalarized solution, forces the partially reflected waves from the PS to mirror off the AdS boundary and re-perturb the PS to give rise to a damped beating pattern which is strikingly similar to echoes from quantum corrected compact objects \cite{Chatzifotis:2020oqr}. The existence of echoes in such AdS-like BHs may lead to interesting interpretations of such a phenomenon on dual holographic descriptions at the boundary, as AdS/CFT suggests (see \cite{Dey:2020lhq,Dey:2020wzm} for a holographic description of echoes on a dual CFT at the near-horizon quantum structure).

Interestingly, even though the wormhole studied here possesses a single barrier effective potential (for non-extremal masses), with its throat located at the peak \cite{Konoplya:2018ala} (similar to that of Bronnikov-Ellis \cite{Bronnikov:1973fh,Ellis:1973yv} and Morris-Thorne \cite{Morris} wormholes), we still observe echoes in the late-time response of the test scalar field due to the existence of the effective AdS boundary. Moreover, if the wormhole is near-extremal or extremal, the response of the probe scalar field exhibits primary and secondary echoes, associated with the AdS boundary and throat potential well, respectively. The concatenation of echoes we observed here are very similar to those found in \cite{Li:2019kwa}, though our setup consists of naturally arising trapping regions, solely depending on the spacetime and perturbation parameters, in contrast to the reflective boundaries placed by hand in \cite{Li:2019kwa}.

Contrary to to the BH case, the echoes found in the wormhole background do not decay with time, but have constant and equal amplitude to that of the initial ringdown. The constancy of the amplitude of echoes is related to the absence of dissipation and may be an indication of the existence of normal oscillation modes, as well as potential instabilities, similar to that found in \cite{Destounis:2019hca}. The introduction of gravitational perturbations (such as the ones considered in \cite{Bronnikov:2004ax,Gonzalez:2008wd,Gonzalez:2009hn,Bronnikov:2012ch}) may lead to an unstable wormhole, due to the presence of phantom matter, which can potentially expand or collapse into a BH \cite{Gonzalez:2008xk,Doroshkevich:2008xm}.

\section*{Acknowledgments}
The authors would like to thank Rodrigo Fontana for helpful discussions.	

%-------BIBLIOGRAPHY--------------

\end{document}